\documentclass[12pt,preprint]{aastex}

\usepackage{epsfig}
\usepackage{lscape}
\usepackage{natbib}
\def\plotone#1{\centering \leavevmode
\includegraphics[clip=, width=.85\columnwidth]{#1}}

\def\plottwo#1#2{\centering \leavevmode
\includegraphics[width=.45\columnwidth]{#1} \hfil
\includegraphics[width=.45\columnwidth]{#2}}

\newcommand{\cN}[1]{\mathcal{N}}
\newcommand{\pn}[1]{\mbox{$(#1)$}}

\def\gsim{\;\rlap{\lower 2.5pt
 \hbox{$\sim$}}\raise 1.5pt\hbox{$>$}\;}
\def\lsim{\;\rlap{\lower 2.5pt
   \hbox{$\sim$}}\raise 1.5pt\hbox{$<$}\;}

\begin{document}


\title{%
Can TiO Explain Thermal Inversions in the Upper Atmospheres of
Irradiated Giant Planets?\\
}

\author{David S. Spiegel\altaffilmark{1}, Katie Silverio\altaffilmark{1},
Adam Burrows\altaffilmark{1}}

\affil{$^1$Department of Astrophysical Sciences, Princeton University, Peyton Hall, Princeton, NJ 08544}

\vspace{0.5\baselineskip}

\email{dsp@astro.princeton.edu, silverio@astro.princeton.edu,
  burrows@astro.princeton.edu}

\begin{abstract}
{\it Spitzer Space Telescope} infrared observations indicate that
several transiting extrasolar giant planets have thermal inversions in
their upper atmospheres.  Above a relative minimum, the temperature
appears to increase with altitude.  Such an inversion probably
requires a species at high altitude that absorbs a significant amount
of incident optical/UV radiation.  Some authors have suggested that
the strong optical absorbers titanium oxide (TiO) and vanadium oxide
(VO) could provide the needed additional opacity, but if regions of
the atmosphere are cold enough for Ti and V to be sequestered into
solids they might rain out and be severely depleted.  With a model of
the vertical distribution of a refractory species in gaseous and
condensed form, we address the question of whether enough TiO (or VO)
could survive aloft in an irradiated planet's atmosphere to produce a
thermal inversion.  We find that it is unlikely that VO could play a
critical role in producing thermal inversions.  Furthermore, we find
that macroscopic mixing is essential to the TiO hypothesis; without
macroscopic mixing, such a heavy species cannot persist in a planet's
upper atmosphere.  The amount of macroscopic mixing that is required
depends on the size of condensed titanium-bearing particles that form
in regions of an atmosphere that are too cold for gaseous TiO to
exist.  We parameterize the macroscopic mixing with the eddy diffusion
coefficient $K_{zz}$ and find, as a function of particle size $a$, the
values that $K_{zz}$ must assume on the highly irradiated planets
HD~209458b, HD~149026b, TrES-4, and OGLE-TR-56b to loft enough
titanium to the upper atmosphere for the TiO hypothesis to be correct.
On these planets, we find that for TiO to be responsible for thermal
inversions $K_{zz}$ must be at least a few times $10^7
\rm~cm^2~s^{-1}$, even for $a = 0.1~\mu$m, and increases to nearly
$10^{11}\rm~cm^2~s^{-1}$ for $a = 10~\mu$m.  Such large values may be
problematic for the TiO hypothesis, but are not impossible.
\end{abstract}


\section{Introduction}
\label{sec:intro}
In the last two decades, we have moved from the first discoveries of
planets beyond our solar system \citep{wolszczan+frail1992,
mayor+queloz1995, marcy+butler1996} to having the ability to frame and
address questions about the structures of distant worlds.  The
remarkable pace of detections has accelerated to the point that there
are now more than $\sim$340 extrasolar planets currently known, of
which more than 55 transit their primaries.\footnote{See the catalog
at http://exoplanet.eu.}

The transiting planets, first discovered by \citet{henry_et_al2000}
and \citet{charbonneau_et_al2000}, constitute a particularly exciting
subsample.  Knowing that their orbits are edge-on breaks the
degeneracy between their masses and orbital inclination angles.
Furthermore, their atmospheres may be probed during both primary
eclipse \citep{seager+sasselov2000, hubbard_et_al2001, brown2001} and
secondary eclipse \citep{sudarsky_et_al2000, sudarsky_et_al2003,
rauscher_et_al2007b, burrows_et_al2008b, fortney_et_al2008}.

Early speculative calculations are now supplemented with a wealth of
observational data that constrain the atmospheres of several
transiting extrasolar giant planets (EGPs).  The strong doublets at
$\sim$5900~\AA~and $\sim$7700~\AA~in neutral sodium (Na) and potassium
(K) contribute significantly to the opacity of planetary atmospheres
through much of the optical range.  The sodium feature has been seen
in the transit spectra of both HD 209458b
\citep{charbonneau_et_al2002, desert_et_al2008} and HD 189733b
\citep{redfield_et_al2008}.  \citet{charbonneau_et_al2002} point out
that the depth of the observed Na feature is less than would be
expected if neutral sodium were present at solar abundance.
\citet{iro_et_al2005} suggest that if, as expected, HD 209458b is in
synchronous rotation, then Na might have condensed into Na$_2$S on the
colder night side, which could in part explain an inferred low
abundance of atomic Na.  Another possible explanation is a high gray
haze \citep{charbonneau_et_al2002, fortney_et_al2003}.

In the simplest picture of an atmosphere's vertical thermal profile,
temperature decreases with altitude (with decreasing pressure).  In
certain cases of high stellar irradiation, however, a chemical species
that is a strong optical or near UV absorber could, if present at high
altitude, lead to a thermal inversion in which the temperature
increases above a relative minimum \citep{hubeny_et_al2003}.  Such an
inversion is well-known in the context of the Earth's stratosphere
\citep{sherwood+dessler2001}, where the upper-atmosphere heating is
caused mainly by UV absorption by ozone.  Infrared observations by the
{\it Spitzer Space Telescope} have suggested that several giant
planets have such thermal inversions, including HD 209458b
\citep{burrows_et_al2007c, knutson_et_al2008b}, HD 149026b
\citep{fortney_et_al2006, burrows_et_al2008b}, TrES-4
\citep{knutson_et_al2009}, XO-1b \citep{machalek_et_al2008}, and,
perhaps, $\upsilon$ Andromeda b \citep{burrows_et_al2008b}.  It is not
known, however, what species is responsible for the additional opacity
high in these planets' atmospheres.

Titanium (Ti) and vanadium (V) both play significant roles in the
spectra of low-mass stellar objects, such as M-dwarfs
\citep{kirkpatrick_et_al1999, martin_et_al1999, lodders2002}.  Their
oxides (TiO and VO) in gaseous form are strong optical absorbers that,
if present at near-solar abundance in the upper atmospheres of highly
irradiated planets, could cause thermal inversions
\citep{hubeny_et_al2003, fortney_et_al2006, fortney_et_al2008,
burrows_et_al2008b}.  It might be difficult, though, to maintain
significant quantities of gaseous TiO/VO aloft.  First, TiO and VO are
significantly heavier than the primary constituent of EGP atmospheres,
molecular hydrogen.  In the battle of molecular diffusion against
gravitational settling, heavier molecules settle more strongly and are
concentrated at deep layers.  Macroscopic mixing processes, such as
turbulent diffusion or large-scale advective motions, are then
required to maintain a high abundance of a heavy species at altitude.

Second, and perhaps more significant, Ti and V might rain out of the
upper atmospheres.  In chemical equilibrium, at low temperatures or
high pressures, Ti and V are found in a variety of condensates that
form solid grains \citep{burrows+sharp1999, lodders2002,
sharp+burrows2007}.  Radiative equilibrium radiative transfer models
of close-in EGP atmospheres \citep{sudarsky_et_al2003,
hubeny_et_al2003, burrows_et_al2007, fortney_et_al2006,
fortney_et_al2008} predict that, in at least some of the planets
inferred to have inversions (e.g., HD 209458b), there is a cooler
region below the inversion in which the temperature and pressure would
cause Ti and V to be in their condensed, solid phases.  Such a region
may aptly be called a ``cold-trap,'' in analogy with the Earth's
equatorial water cold-trap.  Above the Earth's equator, the atmosphere
cools and reaches a relative minimum at the tropopause, the beginning
of the stratospheric thermal inversion.  As a result, even though
water could exist in gaseous form at altitude, most water condenses
and rains out before ever reaching the stratosphere, leaving the
stratosphere quite dry \citep{brewer1949, sherwood+dessler2001}.  If a
similar rain-out process occurs in the Ti/V cold-traps of giant
planets, their upper atmospheres could be similarly depleted of TiO
and VO.

Furthermore, there could be other cold-traps.  Short-period EGPs are
generally presumed to be tidally locked in synchronous rotation
\citep{goldreich+peale1966, spiegel_et_al2007b}.  The large contrast
between the intense stellar forcing at the substellar point and the
minimal heating at the poles and on the night sides of synchronously
rotating planets can lead to large temperature differences and
powerful winds, as has been predicted by atmospheric general
circulation models (GCMs) \citep{showman+guillot2002, cho_et_al2003,
menou_et_al2003, cooper+showman2005, burkert_et_al2005,
langton+laughlin2007, langton+laughlin2008, langton+laughlin2008b,
cho_et_al2008, dobbs-dixon+lin2008, showman_et_al2008,
dobbs-dixon2008, menou+rauscher2008, showman_et_al2008b}.  Transport
of gaseous TiO/VO to the night side of a planet by zonal winds could
lead to the condensation and settling out of these odixes, as
considered by \citet{showman_et_al2008b}.  Though settling might be
slow on orbital timescales (depending on condensate size), if the
night side remains a sink for hundreds of millions of years or more it
could inevitably lead to significant depletion of a planet's upper
atmosphere.

The same types of mixing processes that might loft heavy molecules to
greater altitudes than could molecular diffusion might also help loft
grains or droplets.  If turbulent mixing on a macroscopic scale is
vigorous enough, the condensates of Ti/V in the cold-trap could be
stirred up into the hot, more rarified upper atmosphere, where they
might reform their optically important oxides.

Nevertheless, recent observational evidence suggests that the upper
atmosphere of HD 209458b might indeed be deficient in TiO.
\citet{desert_et_al2008} find that, redward of the Na D lines, the
near-constancy of the planet's transit radius places a limit on the
TiO abundance of $\sim$10$^{-2}-10^{-3}$ solar, an amount that, as we
show in \S\ref{sec:model}, is insufficient to explain the inferred
thermal inversion.  This observation is not proof that TiO is
underabundant in the upper atmosphere of HD 209458b, because transit
spectroscopy probes only atmospheric regions near the terminator.  It
is conceivable that the terminator is depleted of TiO, but the
substellar point is not.

To address this issue, this paper introduces a model for the
abundances of TiO and VO in the upper atmospheres of highly irradiated
EGPs.  We conclude that VO does not contribute significantly to
thermal inversions.  We use our model to estimate the minimum amount
of macroscopic mixing that would be necessary to make the TiO
hypothesis viable on several highly-irradiated planets.  We find that
for HD~209458b, HD~149026b, TrES-4, OGLE-TR-56b, and WASP-12b,
significant macroscopic mixing would be necessary, an amount that it
is not at all clear actually obtains.  Our model predicts that, for
TiO to be present in the upper atmosphere at sufficient quantity to
cause thermal inversion, $K_{zz}$ must be at least a few times $10^7
\rm~cm^2~s^{-1}$ for 0.1-$\mu$m particles, increasing roughly linearly
with condensate particle size.  HD~209458b has the most severe cold
trap of the planets we considered.  There, for particles from
0.1~$\mu$m to 10~$\mu$m in radius, $K_{zz}$ must be nearly $\sim$10$^9
\rm~cm^2~s^{-1}$ to nearly $\sim$10$^{11} \rm~cm^2~s^{-1}$,
respectively.  For larger particles, even greater values of $K_{zz}$
would be required.  WASP-12b has no day-side cold trap in our models;
there, we find that $K_{zz}$ must be $\sim$$2\times
10^7\rm~cm^2~s^{-1}$.

The remainder of this paper is structured as follows: In
\S\ref{sec:model}, we present the model.  We describe how we generate
temperature-pressure profiles and spectra of irradiated expolanets,
and we parameterize the upper atmosphere abundance of a condensing
species in the presence of turbulent and molecular diffusion.
Section~\ref{sec:results} contains the results of our calculations,
and \S\ref{sec:caveats} discusses a few remaining complications.
Finally, in \S\ref{sec:conc}, we summarize our findings.

\section{Modeling the Vertical Distribution of Condensates}
\label{sec:model}
Atmospheres mix their constituents in a variety of ways, at a range of
different spatial scales.  At the micro-level, molecular diffusion
operates; at the macro-level, turbulent mixing and large-scale
advective motions that (including planetary-scale winds) are likely to
be dominant.  How much TiO would have to be mixed through these
various processes, up to the low pressures of the upper atmosphere of
an EGP, in order to cause a thermal inversion?  To address this
question, in \S~\ref{ssec:num_meth} we produce one-dimensional
radiative transfer atmosphere models.  We then, in
\S~\ref{ssec:model}, examine how titanium condensates are likely to be
distributed within these atmosphere models and, therefore, how much
gaseous TiO is actually likely to reach the high parts of the
atmosphere, where it would have to be if it indeed is the needed extra
absorber.

\subsection{The Radiative Model}
\label{ssec:num_meth}
To generate temperature-pressure profiles and spectra for the day and
night sides of several EGPs, we compute model atmospheres with the
code {\tt COOLTLUSTY}, described in \citet{hubeny_et_al2003},
\citet{sudarsky_et_al2003}, \citet{burrows_et_al2006}, and
\citet{burrows_et_al2008b}.  This code is an offshoot of the code {\tt
TLUSTY} \citep{hubeny1988,hubeny+lanz1995}, with chemical abundances
and molecular opacities appropriate to the (relatively) cooler
environments of planetary atmospheres and brown dwarfs
\citep{sharp+burrows2007, burrows+sharp1999, burrows_et_al2001}.
Irradiation in planetary atmosphere models is incorporated using
Kurucz model stellar spectra, interpolated to temperatures and surface
gravities appropriate to exoplanet host stars
\citep{kurucz1979,kurucz1994,kurucz2005}.

In most of our radiative transfer model calculations, we assume
gaseous TiO or VO throughout the entire atmosphere.  We note that
analysis based on this assumption is not strictly self-consistent,
because we use the $T$-$P$ profiles generated from this assumption to
calculate where in fact these gaseous species exist.  Nonetheless,
because our models indicate that gaseous TiO and VO have most of their
radiative influence high in a planet's atmosphere, the model condition
that these species are present in cold-trap regions serves mainly to
avoid discontinuities in opacity versus depth and, therefore, to aid
convergence of the models.  This procedure should not introduce large
errors.

\citet{burrows_et_al2006} propose a formalism for treating the
redistribution of incident stellar flux in a planet's atmosphere in
which the proportion of day-side bolometric stellar flux that is
transported to, and reradiated from, the night side is $P_n$.  This
redistribution parameter plausibly ranges between 0, corresponding to
all stellar flux being instantaneously reradiated, and 0.5,
corresponding to the night side receiving exactly as much stellar
energy as the day side (as a result of advective heat redistribution).
\citet{burrows_et_al2008b} suggest that in many situations $P_n$ is
likely to vary between 0.1 and 0.4.  In this study, we take $P_n =
0.3$, and apply the redistribution between 0.01 and 0.1 bars, for all
models.

\begin{figure}
\plotone{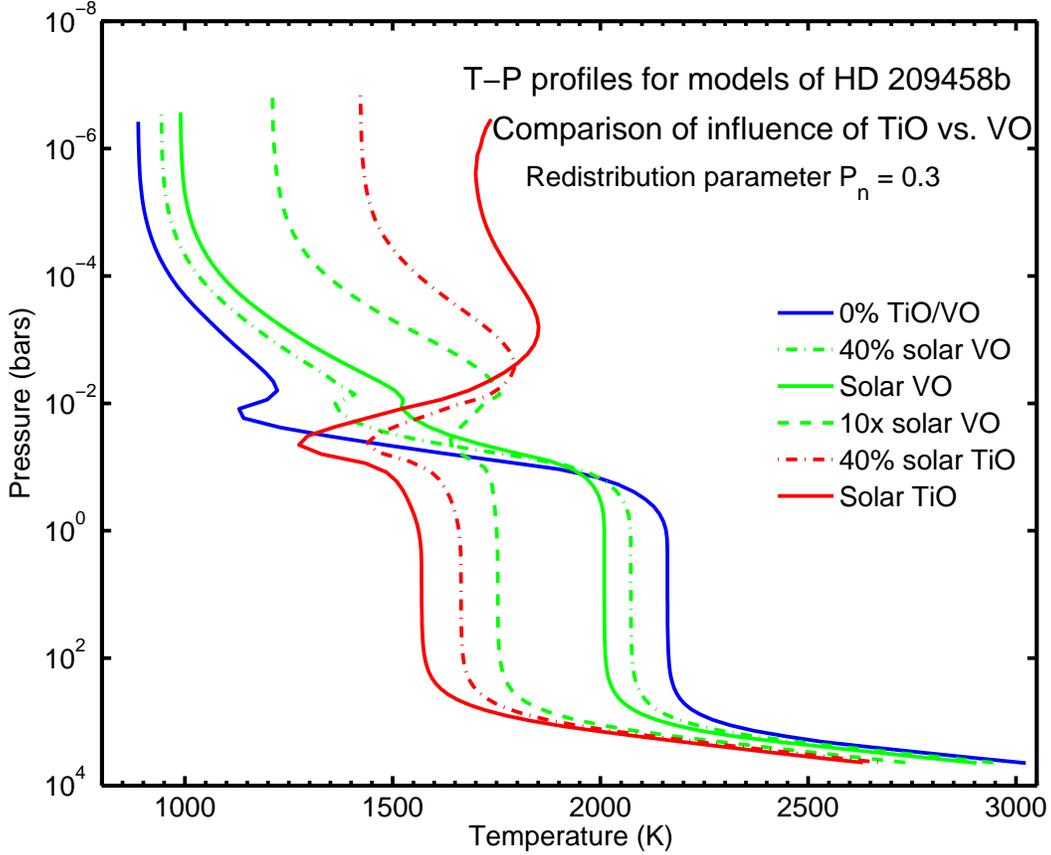}
\caption{Comparison of the influence of TiO and VO.
Temperature-pressure profiles for six models of HD 209458b are
presented, one with no TiO or VO (blue), three with different mixing
ratios of VO, but not TiO (green), and two with TiO, but not VO.  Even
10 times solar abundance of VO has less of an influence on the profile
than just 40\% solar TiO.  This suggests that it is unlikely that VO
plays an important role in determining whether a highly irradiated EGP
has a thermal inversion.  The dimple at $\sim$10$^{-2}$~bars in the
profile for the model with no TiO/VO is caused by the redistribution
to the night side, parameterized by $P_n=0.3$
\citep{burrows_et_al2006}.}
\label{fig:TiO_vs_VO}
\end{figure}

First, we note that it is unlikely that VO plays a crucial determining
role in an EGP's atmospheric structure.  Titanium and vanadium are
both trace elements, but Ti is about $\sim$10 times as abundant as V:
Ti's solar abundance relative to hydrogen is $\sim 10^{-7}$, whereas
V's is $\sim 10^{-8}$ \citep{anders+grevesse1989}.  Furthermore, TiO's
optical opacity is generally greater than that of VO.
Figure~\ref{fig:TiO_vs_VO} presents a comparison of
temperature-pressure profiles for six models of HD 209458b, including
models that contain VO, but not TiO, and models containing TiO, but
not VO.  Although VO does heat the upper atmosphere somewhat, even 10
times solar abundance of VO is insufficient to cause a true thermal
inversion.  In contrast, just 40\% solar abundance of TiO causes a
modest thermal inversion, while solar abundance of TiO causes a
significant inversion in which the upper atmosphere ($\sim$millibar)
is several hundred degrees warmer than the isothermal layer deeper in
($\sim$ 1-100 bars).

\begin{landscape}
\begin{figure}
\plottwo
{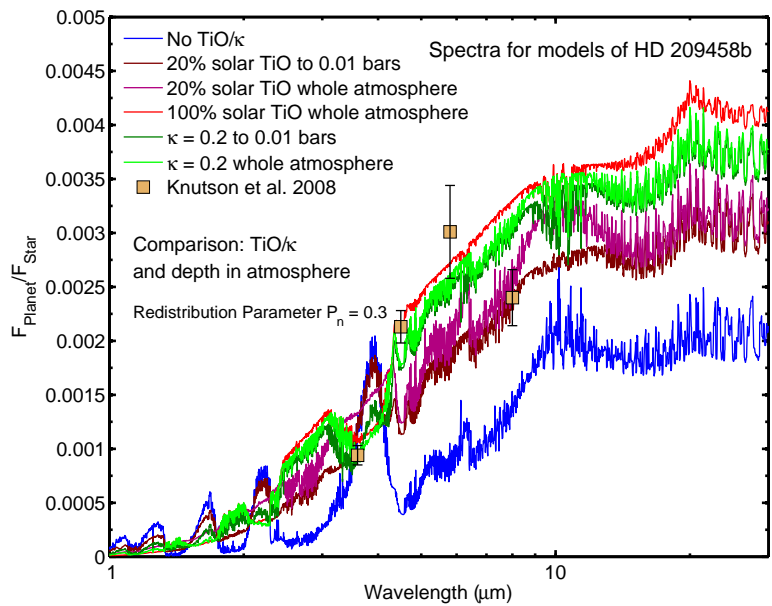}
{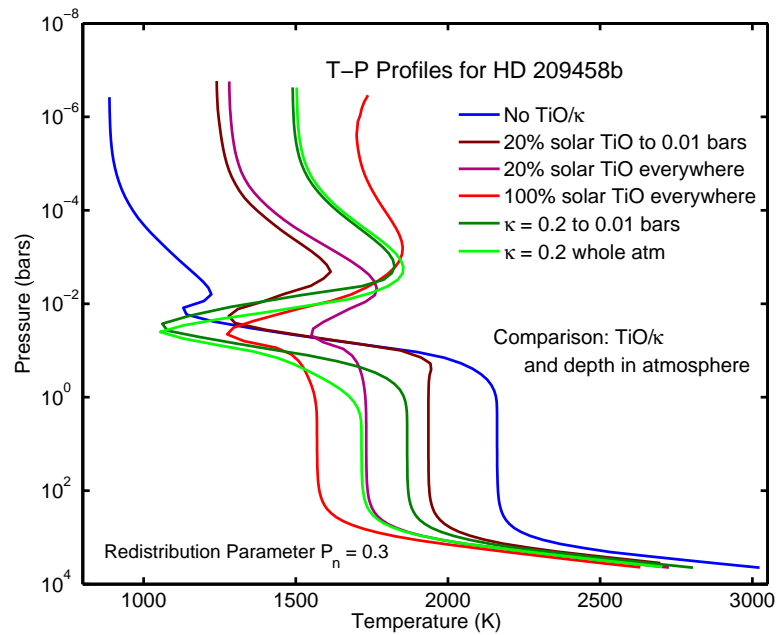}
\caption{Comparison of the effect of model gaseous TiO in the whole
atmosphere versus gaseous TiO only in the upper atmosphere.  This
figure presents spectra (left panel) and temperature-pressure profiles
(right panel) for six models of HD 209458b.  Superposed on the
spectrum plot are the IRAC data points from
\citet{knutson_et_al2008b}.  There are two models with 20\% solar TiO,
one of which has this mixing ratio throughout the whole atmosphere and
the other with TiO added only above 0.01 bars. There are two models
with an absorber, that is gray between $3\times 10^{14}$ and $7\times
10^{14}$~Hz, whose opacity is $\kappa = 0.2 {\rm~cm^2~g^{-1}}$ (the
$\kappa_e$ of \citealt{burrows_et_al2007}); one has the absorber
throughout the whole atmosphere and one has the absorber only above
0.01 bars.  Finally, there is both a model with no TiO and a model
with solar abundance of TiO throughout the atmosphere.  See
\S~\ref{ssec:num_meth} for a discussion.}
\label{fig:HD209_spectra_TiO_kappa}
\end{figure}
\end{landscape}

We, therefore, neglect the contributions of VO, and investigate what
errors in spectra and temperature-pressure profiles are likely to be
generated by the model assumption that gaseous TiO exists throughout
the atmosphere.  Figure~\ref{fig:HD209_spectra_TiO_kappa} portrays
spectra (left panel) and temperature-pressure profiles (right panel)
for six models of HD 209458b.  Superposed on the spectrum plot are the
four data points for this planet measured by
\citet{knutson_et_al2008b} using the {\it Spitzer} InfraRed Array
Camera (IRAC).  One of the models in this figure is a base case that
has no additional upper atmosphere absorber (blue curve); this model
does not produce a thermal inversion and fails to match the
\citet{knutson_et_al2008b} data.  Another model has solar abundance of
TiO throughout the atmosphere (red curve); this one has the largest
thermal inversion and matches the \citet{knutson_et_al2008b} data the
best. The remainder of the models in this figure demonstrate that our
assumption that the extra absorber is present throughout the
atmosphere does not cause large errors either in predicted emergent
spectra or in temperature-pressure profiles.  Included on
Fig.~\ref{fig:HD209_spectra_TiO_kappa} is a pair of models with 20\%
solar TiO.  One model has this species present from the top of the
atmosphere down to 0.01 bars, and the other has it throughout the
atmosphere.  Finally, there is a pair of models with a gray
absorber\footnote{The absorber is not truly gray, but rather has an
opacity that is a top-hat function between $3\times 10^{14}$ and
$7\times 10^{14}$~Hz.} of opacity $\kappa = 0.2 \rm~cm^2~g^{-1}$,
similar to the $\kappa_e$ of \citet{burrows_et_al2008b}.  Again, one
model has this absorber from the top of the atmosphere to 0.01 bars
and the other has it throughout the atmosphere.  The differences in
spectra and profiles between the models with absorbers only in the
upper atmosphere and those with absorbers throughout the atmosphere
are minor enough that for the purposes of this simple study we proceed
with models that have the absorber everywhere.

\begin{landscape}
\begin{figure}
\plottwo
{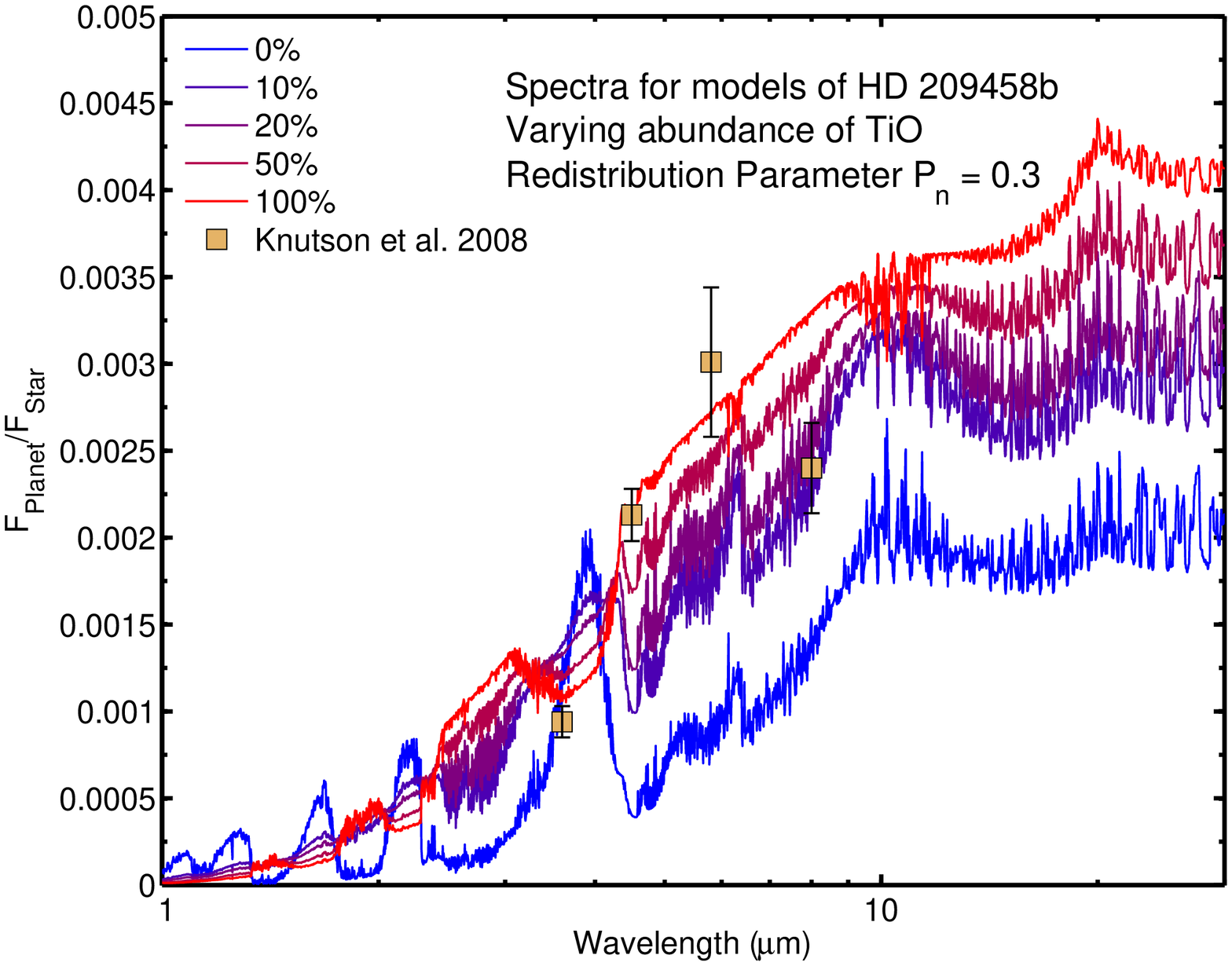}
{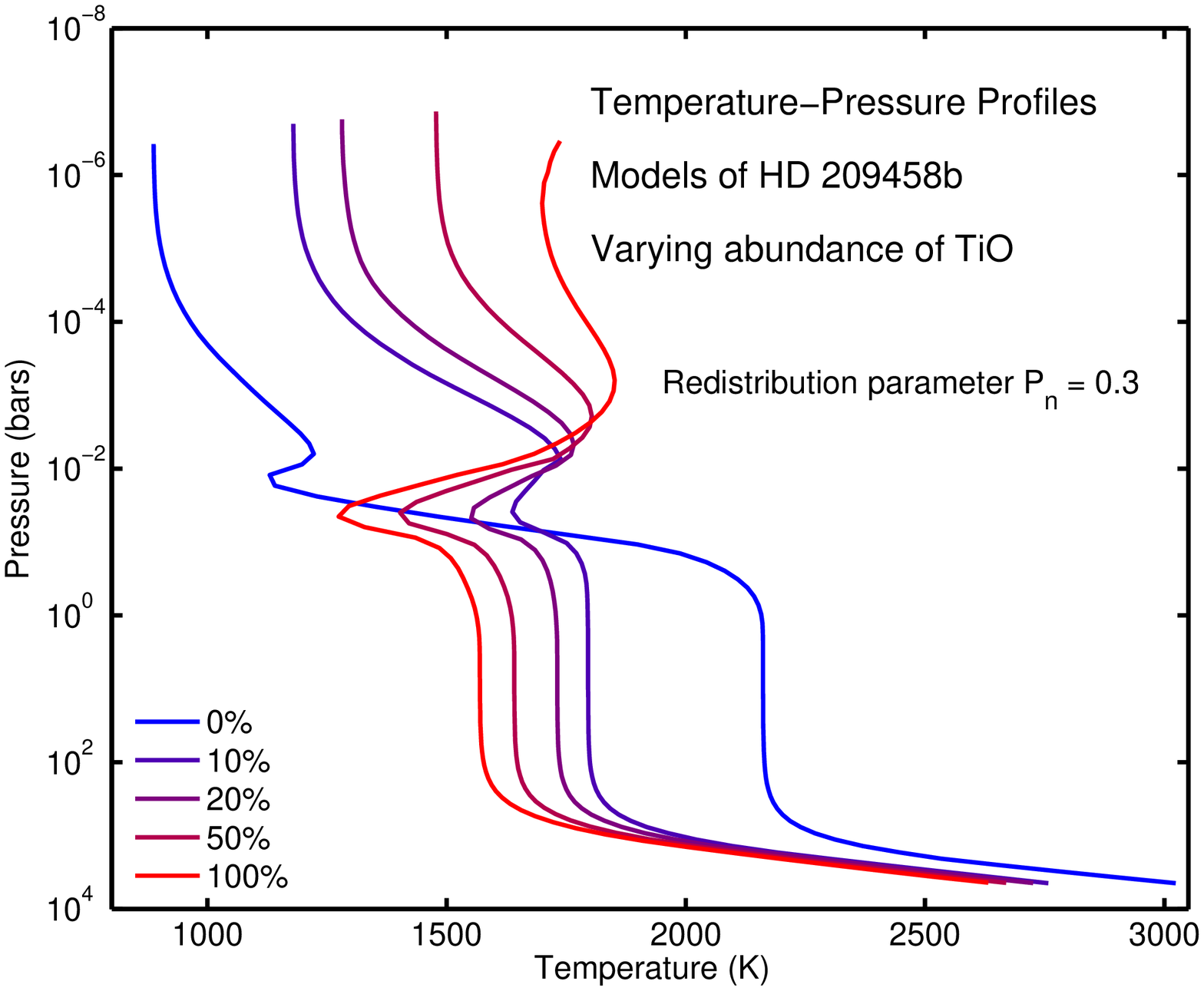}
\caption{Comparison of different mixing ratios of TiO.  This figure
shows spectra (left panel) and temperature-pressure profiles (right
panel) for five models of HD 209458b: with no TiO, and with 10\%,
20\%, 50\%, and 100\% of solar abundance of TiO.  Gaseous TiO is
assumed throughout the atmosphere, and the redistribution parameter
$P_n = 0.3$ is used for all models.  The spectrum plot superposes IRAC
data points from \citet{knutson_et_al2008b}.  As the abundance of TiO
increases, the upper atmosphere temperature increases, since it
absorbs a greater fraction of incident stellar energy, while the lower
atmosphere cools because less of the stellar flux penetrates to depth.
The models with 50\% and 100\% solar abundance TiO have significant
thermal inversions in their upper atmospheres; the models with less
TiO do not.  It is most significant that higher mixing ratios of TiO
cause greater planet-star flux ratios over most of the wavelength
range.  Models with 50\% and 100\% solar abundance of TiO are decent
matches to the IRAC 1 ($\sim$3.6~$\mu$m), IRAC 2 ($\sim$4.5~$\mu$m),
and IRAC 3 ($\sim$5.8~$\mu$m) points, though they do fail to match the
IRAC 4 ($\sim$8.0~$\mu$m) point.  However, the 0\%, 10\%, and 20\%
solar TiO models, which lack thermal inversions, entirely fail to
match the IRAC data.}
\label{fig:HD209_spectra_TiO}
\end{figure}
\end{landscape}

Finally, we consider several VO-free models of HD 209458b in order to
estimate what mixing ratio of TiO would be required to sustain an
upper atmosphere inversion that produces spectra roughly consistent
with measured data.  Figure~\ref{fig:HD209_spectra_TiO} presents
spectra (left panel) and temperature-pressure profiles (right panel)
for five models of HD 209458b, one with no TiO, and others with 10\%,
20\%, 50\%, and 100\% solar abundance of TiO.  Superposed on the
spectrum plot are the four IRAC points from
\citet{knutson_et_al2008b}.  The models with 50\% and 100\% solar
abundance of TiO have thermal inversions in the upper atmosphere and
are consistent with the IRAC 1 ($\sim$3.6~$\mu$m), 2
($\sim$4.5~$\mu$m), and 3 ($\sim$5.8~$\mu$m) points.  The fact that
none of these models matches all of the observed data shows that our
theoretical understanding of radiation from exoplanet atmospheres is
incomplete.  Nevertheless, one crucial lesson from the data of
\citet{knutson_et_al2008} is that since the IRAC 2 point is
significantly higher than the IRAC 1 point, and since the photosphere
of IRAC 2 is at greater altitude than that of IRAC 1
\citep{burrows_et_al2008b}, the planet would seem ineluctably to have
a thermal inversion, other theoretical uncertainties notwithstanding.
Fig.~\ref{fig:HD209_spectra_TiO} suggests that if TiO is the extra
absorber its mixing ratio in the upper atmosphere ought to be no less
than $\sim$50\% of the corresponding solar ratio.

\subsection{Modeling Mixing Ratio vs. Altitude}
\label{ssec:model}
As Fig.~\ref{fig:HD209_spectra_TiO} shows, any cold-trap regions
cannot deplete the atmosphere of Ti too significantly without leaving
TiO insufficiently abundant to produce the inferred thermal
inversions.  How can we estimate how much the cold-trap region
depletes the upper atmosphere of TiO?  In this section, we describe
our model for determining the atmospheric profiles of TiO and the
amount of depletion in a turbulent cold-trap region.  To do so, we
introduce the turbulent diffusion coefficient $K_{zz}$
\citep{colegrove_et_al1965, lewis+fegley1984, noll_et_al1988,
drossart_et_al1990, rodrigo_et_al1990}.  $K_{zz}$ parameterizes, in a
single number, a variety of processes (including turbulence and other
forms of macroscopic mixing) that act for each species to equalize the
number density (or partial pressure) at different spatial locations.

\begin{figure}
\plotone{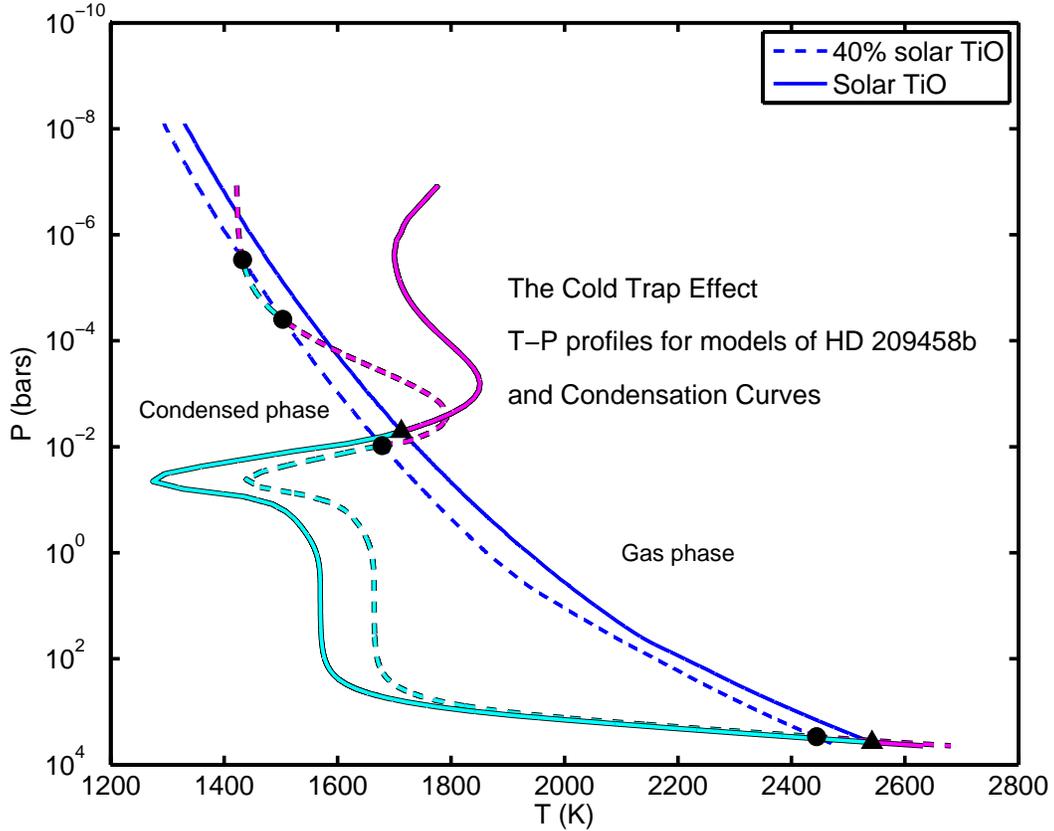}
\caption{The TiO cold-trap.  Condensation curves (blue) and
temperature-presure profiles of models of HD 209458b (cyan and
magenta) are shown for two mixing ratios of TiO -- 40\% solar (dashed)
and 100\% solar (solid). The profiles are calculated assuming TiO is
present in the entire atmosphere at the specified mixing ratio.
Magenta (cyan) parts of the profiles indicate where the atmosphere is
warmer (colder) than the corresponding condensation curve and,
therefore, where titanium is in the gaseous TiO (condensed)
phase. Filled circles (triangles) mark the intersections of the 40\%
(100\%) solar mixing ratio profile with the corresponding condensation
curve.  Notice that, at 40\% solar titanium, the atmosphere of HD
209458b might have two titanium cold-traps, whereas at 100\% solar it
has just one.}
\label{fig:cold_trap}
\end{figure}

We start by identifying the cold-trap region(s).  On the day side,
there are in general three regions: \pn{i} the hot, isentropic
convection zone, which contains gaseous TiO; \pn{ii} the cold trap,
which contains titanium condensates; and, \pn{iii} the hot, thermally
inverted upper atmosphere, in which TiO is gaseous.  Depending on a
planet's temperature-pressure profile, it might also have no
cold-traps.  The phase of titanium compounds as a function of altitude
in a planet's atmosphere is found by comparing the planet's
temperature-pressure profile with the titanium condensation curve for
the corresponding metallicity.  TiO is gaseous where the atmosphere is
hotter than the condensation curve and titanium is in condensed form
where the atmosphere is cooler than the condensation curve.  These
condensation curves vary with metallicity.  At a given pressure,
higher metallicity causes the condensation curve to be at a higher
temperature.  Figure~\ref{fig:cold_trap} illustrates, in the context
of HD 209458b, how the comparison of profiles to condensation curves
yields the location of cold-trap regions.  Models of this planet with
40\% solar TiO and with 100\% solar TiO are presented, along with the
corresponding condensation curves \citep{sharp+burrows2007}.  The
points of intersection are demarcated, and cold-trap regions are
indicated by cyan regions of the profile curves.  For example, on HD
209458b in the two profiles shown the largest cold-trap region extends
from roughly $3 \times 10^3$ bars to $10^{-2}$ bars.

In order to understand how much Ti reaches the upper atmosphere of an
EGP, we start by considering the vertical distribution of different
molecular species in an atmosphere with a variety of chemical
constituents.  For a particular molecular species $i$ (where $i$ could
be TiO, H$_2$, or any other atmospheric constituent), the vertical
abundance profile $n_i[z]$ varies with time as $\partial n_i /
\partial t = - \nabla \cdot \vec{F}_i ~(\rm +~ sources_i ~ - ~
sinks_i)$, where $n_i[z]$ is the number density of species $i$ as a
function of altitude $z$, and $\vec{F}_i$ is the flux of species $i$.
If we ignore sources and sinks, and express the flux as a combination
of molecular and turbulent diffusion with gravitational settling, as
per \citet{brewer1949} and \citet{chamberlain+hunten1987}, then we may
write the rate of change of the vertical distribution of species $i$
as follows:
\begin{eqnarray}
\nonumber \frac{\partial n_i[z]}{\partial t} & = & \frac{\partial}{\partial z} \left\{ D_i \left( \frac{\partial n_i}{\partial z} + \left( \frac{\partial \ln T}{\partial z} + \frac{m_i g}{kT} \right)n_i \right) \right. \\
\label{eq:diffusion_eq} & & \left. + K_{zz} \left( \frac{\partial n_i}{\partial z} + \left( \frac{\partial \ln T}{\partial z} + \frac{\mu m_p g}{kT} \right) n_i \right) \right\}.
\end{eqnarray}
In this equation, $D_i$ is the coefficient of molecular diffusion,
$m_i$ is the molecular weight, $\mu m_p$ is the product of the mean
molecular weight ($\mu$) and the mass of a proton ($m_p$), $g$ is the
gravitational acceleration (assumed to be constant), $k$ is
Boltzmann's constant, $T$ is the temperature, and $K_{zz}$ is the
coefficient of turbulent diffusion in the vertical direction.

The timescale for achieving steady state in a diffusion problem with
eddy diffusion coefficient $K_{zz}$ is given by the square of a
characteristic length divided by $K_{zz}$ \citep{griffith+yelle1999}:
\begin{equation}
\tau_{\rm ss} \sim L^2/K_{zz} \, .
\label{eq:time_ss}
\end{equation}
A characteristic vertical length scale is the pressure scale height
\begin{equation}
H_P \equiv \frac{k T}{\mu m_p g} \, ,
\label{eq:Hp}
\end{equation}
which is not more than $\sim$10$^8$~cm (1000~km) for EGPs.  $K_{zz}$
is not well constrained.  A variety of estimates of $K_{zz}$ for
Jupiter's atmosphere place it in the range $10^2 - 10^9 \rm
~cm^2~s^{-1}$ \citep{bezard_et_al2002, fegley+lodders1994,
ackerman+marley2001, benjaffel_et_al2007}.  Estimates in the context
of brown dwarfs place it in the range $10^2 - 10^6 \rm ~cm^2~s^{-1}$
\citep{saumon_et_al2006, saumon_et_al2007, griffith+yelle1999,
hubeny+burrows2007}.  If highly irradiated EGPs are at least as
vigorously mixed as Jupiter and brown dwarfs are thought to be, then
we may find an upper bound for $\tau_{\rm ss}$ by setting $K_{zz}
\gsim 10^2 \rm~cm^2~s^{-1}$ in eq.~(\ref{eq:time_ss}):
\begin{equation}
\tau_{\rm ss} \lsim 10^{14}~\rm s \sim 3~\rm Myr \ll \tau_{\rm planet~age} \, ,
\label{eq:time_ss_limit}
\end{equation}
where $\tau_{\rm planet~age}$ is the planet's age.  This indicates
that, in much less than a planet's age (hundreds of megayears to a
gigayear or more), the atmosphere has reached a steady state and
$\partial n_i/\partial t = 0$.  We further assume that the mean net
fluxes are zero in the steady state.  Equation~(\ref{eq:diffusion_eq})
may, therefore, be rewritten as follows:
\begin{equation}
D_i \left( \frac{\partial P_i}{\partial z} + \frac{m_i g}{k T} P_i \right) + K_{zz} \left( \frac{\partial P_i}{\partial z} + \frac{\mu m_p g}{k T} P_i \right) = 0 \, .
\label{eq:brewer_eq}
\end{equation}
Here, the $n_i$ of eq.~(\ref{eq:diffusion_eq}) has been replaced by
the partial pressure of species $i$, $P_i$, in accordance with the
ideal gas law.

The general solution to this equation is
\begin{equation}
P_i[z] = P_i[z_0] \exp \left[ - \int_{z_0}^z d\zeta \frac{(m_i/\mu m_p) D_i + K_{zz}}{H_P[\zeta](D_i + K_{zz})} \right],
\label{eq:brewer_eq_sol}
\end{equation}
where $z_0$ is an arbitrary reference height.  In general, $T$, $\mu$,
$D_i$, and $K_{zz}$ are functions of $z$.  Notice that
eq.~(\ref{eq:brewer_eq_sol}) indicates that if vertical mixing is
vigorous ($K_{zz} \gg D_i$), all species will have the same vertical
scale height $H_P$.  Conversely, in the absence of macroscopic mixing
($K_{zz} \ll D_i$), gravitational settling of heavier molecules will
cause each species to have its own pressure scale height $H_{P_i}
\approx H_P/(m_i/\mu m_p)$.

Since the total pressure varies as
\begin{equation}
p[z] = p[z_0] \exp \left[ - \int_{z_0}^z d\zeta/H_P[\zeta] \right],
\label{eq:backg_p}
\end{equation}
the mixing ratio of species $i$, $Y_i \equiv P_i/P$, may be expressed
as
\begin{eqnarray}
\nonumber Y_i[z] & = & Y_i[z_0] \exp \left[ - \int_{z_0}^z d\zeta \left\{ \frac{(m_i/\mu m_p) D_i + K_{zz}}{H_P[\zeta](D_i + K_{zz})} - \frac{1}{H_P[\zeta]} \right\} \right] \\
\label{eq:Yi} & = & Y_i[z_0] \exp \left[ - \int_{z_0}^z \frac{d\zeta}{H_P[\zeta]} \frac{\left\{ (m_i/\mu m_p) - 1 \right\} D_i}{D_i + K_{zz}} \right].
\end{eqnarray}
Equation~(\ref{eq:Yi}) has the same general form as
eq.~(\ref{eq:brewer_eq_sol}), which prompts us to define a ``scale
height of the mixing ratio'' $H_{Y_i}$ as follows:
\begin{equation}
H_{Y_i}[z] \equiv H_P[z] \frac{D_i + K_{zz}}{D_i \left\{ (m_i/\mu m_p) -1 \right\}} \, .
\label{eq:H_Yi}
\end{equation}

We consider the behavior of this model in cases where $K_{zz}$ is
large and where it is small relative to $D_i$.  If $K_{zz} \gg D_i$,
then $H_{Y_i} \gg H_P$, which indicates that the mixing ratio of
species $i$ remains essentially constant with height, as is expected
with vigorous mixing.  If $K_{zz} \ll D_i$, the scale height of the
mixing ratio $H_{Y_i} \approx H_P/\left\{ (m_i/\mu m_p) -1 \right\}$,
which indicates that the $e$-folding height for the mixing ratio of
each species is roughly inversely proportional to its molecular
weight, which we take to be 2.3 in all models.\footnote{In the limit
of zero macroscopic mixing, even helium will separate from molecular
hydrogen.  In this case, $Y_i$ should be replaced by the mixing ratio
of species $i$ relative to H$_2$, the background pressure $P_i$ should
be replaced by the pressure of molecular hydrogen, $P_{\rm H_2}$, and
$\mu m_p$ should be replaced by $2 m_p$.}

Presumably, the deep interior of an EGP is well-mixed by convection,
with $Y_i$ constant with $z$ for all species $i$.  In the atmosphere,
a useful metric of the abundance of a species is its mixing ratio
relative to its mixing ratio at depth:
\begin{equation}
f[z] \equiv \frac{Y_i[z]}{Y_i[0]},
\label{eq:f}
\end{equation}
where $z=0$ is taken to be the radiative-convective boundary.

\subsubsection{Vertical Distribution of Gaseous Species}
\label{sssec:gas}
In regions where TiO is gaseous, how does its mixing ratio vary with
altitude?  In accordance with \citet{kittel1969} we estimate the
diffusivity of TiO through molecular hydrogen as
\begin{equation}
D_{\rm TiO} \sim \frac{\lambda \langle v \rangle}{3} \, ,
\label{eq:D_estimate}
\end{equation}
where $\lambda$ is the mean free path and $\langle v \rangle$ is the
mean velocity.  We estimate the mean free path as $\lambda \sim 1/(n
\sigma)$, where $\sigma$ is the cross-section for collision.  Taking
as a typical cross-section $\sigma \sim \pi (1 \rm~\AA)^2$, and as a
typical speed $\langle v \rangle \sim (8k T / 3m_{\rm TiO})^{1/2}$, we
estimate $D_{\rm TiO} \sim 90 \rm~cm^2~s^{-1}$ at 1500~K and 1~bar.
Similarly, this argument suggests that at kilobar pressures $D_{\rm
TiO}$ is in the range of $\sim$10$^{-1} \rm~cm^2~s^{-1}$, whereas at
millibar pressures, it is in the range of $\sim$10$^5
\rm~cm^2~s^{-1}$.  This simple estimate predicts comparable molecular
diffusion values to those calculated in \citet{rodrigo_et_al1990} in
the context of the Earth's thermosphere.

We now consider several possible cases of relative magnitudes of
molecular and turbulent diffusion:
\begin{enumerate}
\item
If there is no turbulent mixing, or if $K_{zz} \ll D_{\rm TiO}$,
eq.~(\ref{eq:H_Yi}) implies that $H_{Y_{\rm TiO}} \approx H_P / 27$.
On HD 209458b, $H_P \sim 500-600 \rm~km$.  As a result, if $K_{zz} \ll
D_{\rm TiO}$, the scale height of the mixing ratio $H_{Y_{\rm TiO}}
\sim 20$~km.  In the $\sim$14 pressure scale heights from a kilobar to
a millibar, the mixing ratio would drop by $\sim$370 $e$-foldings.  In
short, without macroscopic mixing, $f_{\rm TiO} = 0$ at the top of the
atmosphere.
\item
At the opposite extreme, if $K_{zz} \ggg D_{\rm TiO}$, the mixing
ratio of even a heavy molecule remains essentially constant so long as
the species is in the gas phase.  If, for example, $K_{zz} =
10^4(D_{\rm TiO})$, eq.~(\ref{eq:H_Yi}) implies that $H_{Y_{\rm TiO}}
\approx 370 H_P$.  In this case, in a vertical distance of $\sim$14
pressure scale heights the mixing ratio of TiO would decrease by only
$\sim$4\%.
\item
Finally, consider a regime wherein $K_{zz}$ is somewhat, but not
overwhelmingly, larger than $D_{\rm TiO}$.  If this is the case,
$f_{\rm TiO}$ might be neither $\sim$0 nor $\sim$1, but rather an
interesting intermediate value.  Indeed, since the mean free path
varies inversely with number density -- and, therefore, $D$ varies
roughly with $\sim p^{-1}$ -- it seems likely that there are some
regions of an EGP's atmosphere that are in this regime.  For example,
if $K_{zz} = 10 D_{\rm TiO}$ at some altitude, then $H_{Y_{\rm TiO}}
\sim 0.41 H_P$.  This would imply that the mixing ratio drops off with
altitude more precipitously than pressure does, but not dramatically
more.  If, at another altitude, $K_{zz} = 100 D_{\rm TiO}$, then
$H_{Y_{\rm TiO}} \sim 3.7 H_P$.  This would imply that the mixing
ratio decreases with altitude more gradually than pressure does.
\end{enumerate}
In the general case, where TiO is in the gas phase $f_{\rm TiO}$
varies with altitude according to the following equation:
\begin{equation}
\ln[f_i]  =   - \int_{z_0}^z \frac{d\zeta}{H_P[\zeta]} \frac{\left\{ (m_i/\mu m_p) - 1 \right\} D_i}{D_i + K_{zz}} \, .
\label{eq:f_gas}
\end{equation}

\subsubsection{Vertical Distribution of Condensed Species}
\label{sssec:cond}
In the cold-trap region, titanium is in condensate form.  Details of
the chemistry by which TiO condenses into a variety of Ti-bearing
compounds can be found in \citet{lodders2002}.  Because of the
complicated nature of the condensation chemistry, it is difficult to
specify exactly which condensates form and in what quantity.
Therefore, we will simply use the subscript $c_i$ to denote the
condensates of species $i$.  We assume that the condensates will form
roughly spherical condensates of radius $a$.  Lacking a
first-principles theory for the process of condensate formation, we
cannot predict the size-distribution or the modal value of particle
radii.\footnote{Nucleation theory \citep{draine+salpeter1977,
draine1981, cooper_et_al2003} might at some point provide a method of
estimating particle size, but the community is currently far from an
ab initio theory of condensate sizes in a planetary atmosphere.}
Instead, we use a wide range of values of $a$ that spans several
orders of magnitude: $0.1~\mu {\rm m} \leq a \leq 100~\mu {\rm m}$. To
model the distribution of these condensates, we start with an
expression analogous to eq.~(\ref{eq:diffusion_eq}), only for the
condensed form of species $i$, taking $n_{c_i}$ to be the number
density of condensates of species $i$.  If we consider the cloud of
condensates to be a gas, it is an extremely rarified one.  To a good
approximation, it should, therefore, behave as an ideal gas.  The
``partial pressure'' of condensates of species $i$ is then
approximately given by the ideal gas law: $P_{c_i} \approx n_{c_i} k
T$.

The argument that the atmosphere has achieved steady state applies
here as well. Therefore, we proceed in the same manner as before, by
setting the vertical flux of condensates equal to zero and solving for
$P_{c_i}$.  The solution is given by an expression identical to
eq.~(\ref{eq:brewer_eq_sol}), except with $m_i$ and $D_i$ replaced by
$m_{c_i}$ and $D_{c_i}$.

The background pressure is again given by eq.~(\ref{eq:backg_p}).  If
$N_{c_i} \equiv m_{c_i}/m_i$ is the number of atoms/molecules of
species $i$ per condensates, then, following eq.~(\ref{eq:Yi}), the
mixing ratio in a cold-trap region may be written
\begin{eqnarray}
\nonumber Y_{c_i} & \equiv & N_{c_i} \frac{P_{c_i}}{P} \\
\label{eq:Yci} & = & Y_i[0] \exp \left[ - \int_{0}^z \frac{d\zeta}{H_P[\zeta]} \frac{\left\{ (m_{c_i}/\mu m_p) - 1 \right\} D_{c_i}}{D_{c_i} + K_{zz}} \right].
\end{eqnarray}
The ratio of the mixing ratio $f_i$ of species $i$ to its value at
depth may be written,
\begin{equation}
\label{eq:fc} f[z] = \frac{Y_{c_i}}{Y[0]} \, .
\end{equation}

Equation~(\ref{eq:Yci}) may be simplified by making use of the low
Reynolds number expression for terminal velocity, appropriate for an
EGP's atmosphere \citep{ackerman+marley2001, showman_et_al2008b}.  In
this regime, the viscous drag force, balanced by gravity, is $F_v =
6\pi \eta a v$, where $\eta$ is the dynamic viscosity, $a$ the
particle size, and $v$ the speed of the particle through the viscous
medium.  This expression implies a terminal velocity of
\begin{equation}
{v_f}_{\rm Stokes} = \frac{2}{9} \frac{a^2 \rho_{c_i} g}{\eta}.
\label{eq:vf}
\end{equation}
We estimate the molecular diffusivity of condensates $D_{c_i}$ using
the Stokes-Einstein relation:
\begin{equation}
D_{\rm Stokes} = \frac{k T}{6 \pi \eta a} \, .
\label{eq:stokes-einstein}
\end{equation}
Equation~(\ref{eq:stokes-einstein}) is valid in the low-Reynolds
number regime only when the mean free path of the molecules
constituting the fluid is small compared to the size of the falling
body.  When this ratio, called the Knudsen number ($N_{\rm Kn} \equiv
\lambda/a$; \citealt{knudsen1911}) is not small, a correction is
needed.  The analysis of small particles moving through highly
rarified media has a long history, dating back to early theoretical
work by \citet{cunningham1910} and contemporaneous experimental work
\citep{millikan1913,millikan1923}.  The form of the
Cunningham-Millikan-Davies ``slip factor correction''
\citep{davies1945} has been updated somewhat over the years
\citep{el-fandy1953,baines_et_al1965}.  We adopt the value from
\citet{li+wang2003}:
\begin{equation}
\beta = 1 + N_{\rm Kn}(1.256 + 0.4 \exp[-1.1/N_{\rm Kn}]) \, .
\label{eq:beta}
\end{equation}
In the nonzero Knudsen number regime, the true terminal velocity and
diffusivity are increased over the Stokes values (eqs.~\ref{eq:vf} and
\ref{eq:stokes-einstein}) by a factor of $\beta$:
\begin{eqnarray}
\label{eq:vf_corr} v_f & = & \beta {v_f}_{\rm Stokes} \\
\label{eq:D_corr}  D   & = & \beta D_{\rm Stokes} \, .
\end{eqnarray}

Substituting eq.~(\ref{eq:vf_corr}) into eq.~(\ref{eq:D_corr}), and
noting that $m_{c_i} = (4/3)\pi (a_{c_i})^3 \rho_{c_i}$, we have
\begin{eqnarray}
\nonumber (m_{c_i}/\mu m_p)D_{c_i} & = & v_f k T / (\mu m_p g) \\
 & = & \beta {v_f}_{\rm Stokes} H_P.
\label{eq:mcDc}
\end{eqnarray}

When eq.~(\ref{eq:mcDc}) is substituted into the expression for
$H_{Y_{c_i}}$, the result is
\begin{equation}
H_{Y_{c_i}}^*[z] = H_P[z] \frac{D_{c_i} + K_{zz}}{\beta {v_f}_{\rm Stokes} H_P (1 - \mu m_p / m_{c_i})} = \frac{D_{c_i} + K_{zz}}{\beta {v_f}_{\rm Stokes} (1 - \mu m_p / m_{c_i})} \, .
\label{eq:Hi*2}
\end{equation}
Equation~(\ref{eq:D_corr}) implies that $D_{c_i} = 4.4\times 10^{-7}
{\rm~cm^2~s^{-1}} \beta (a/1{\mu \rm m})^{-1} (T/1500~\rm K)
(\eta/2.5\times 10^{-4} {\rm~g~cm^{-1}})$. If $K_{zz} \gg D_{c_i}$ and
$m_{c_i} \gg \mu m_p$, then
\begin{equation}
H_{Y_{c_i}}^*[z] \approx \frac{K_{zz}}{\beta {v_f}_{\rm Stokes}} \, .
\label{eq:Hi*3}
\end{equation}
This implies that, at the top of a cold-trap of vertical extent
$z-z_0$,
\begin{equation}
\ln[f_i]  =   - \int_{z_0}^z \frac{d\zeta}{K_{zz}/(\beta {v_f}_{\rm Stokes})} \, .
\label{eq:f2}
\end{equation}

\section{Results}
\label{sec:results}
\begin{landscape}
\begin{figure}
\plottwo
{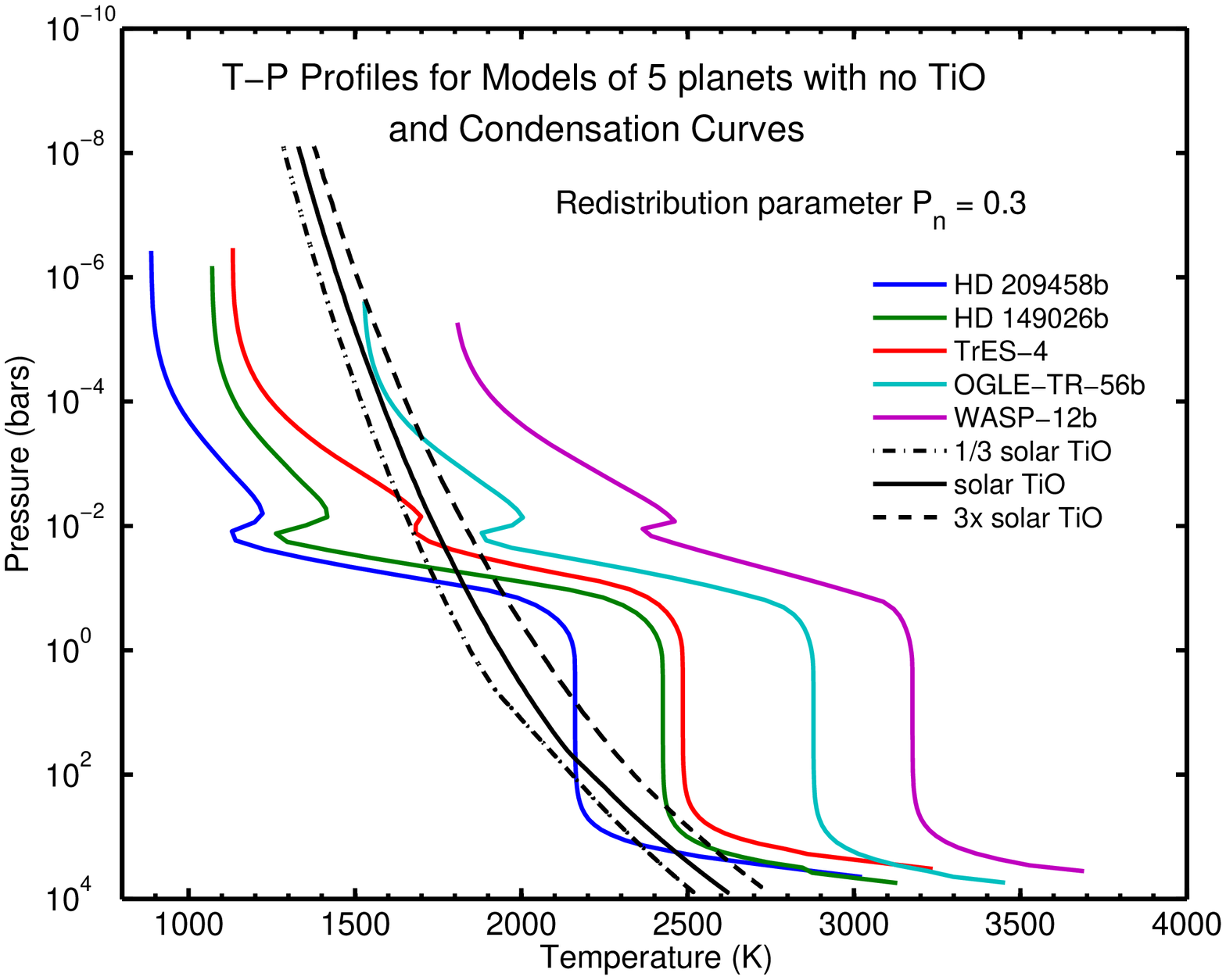}
{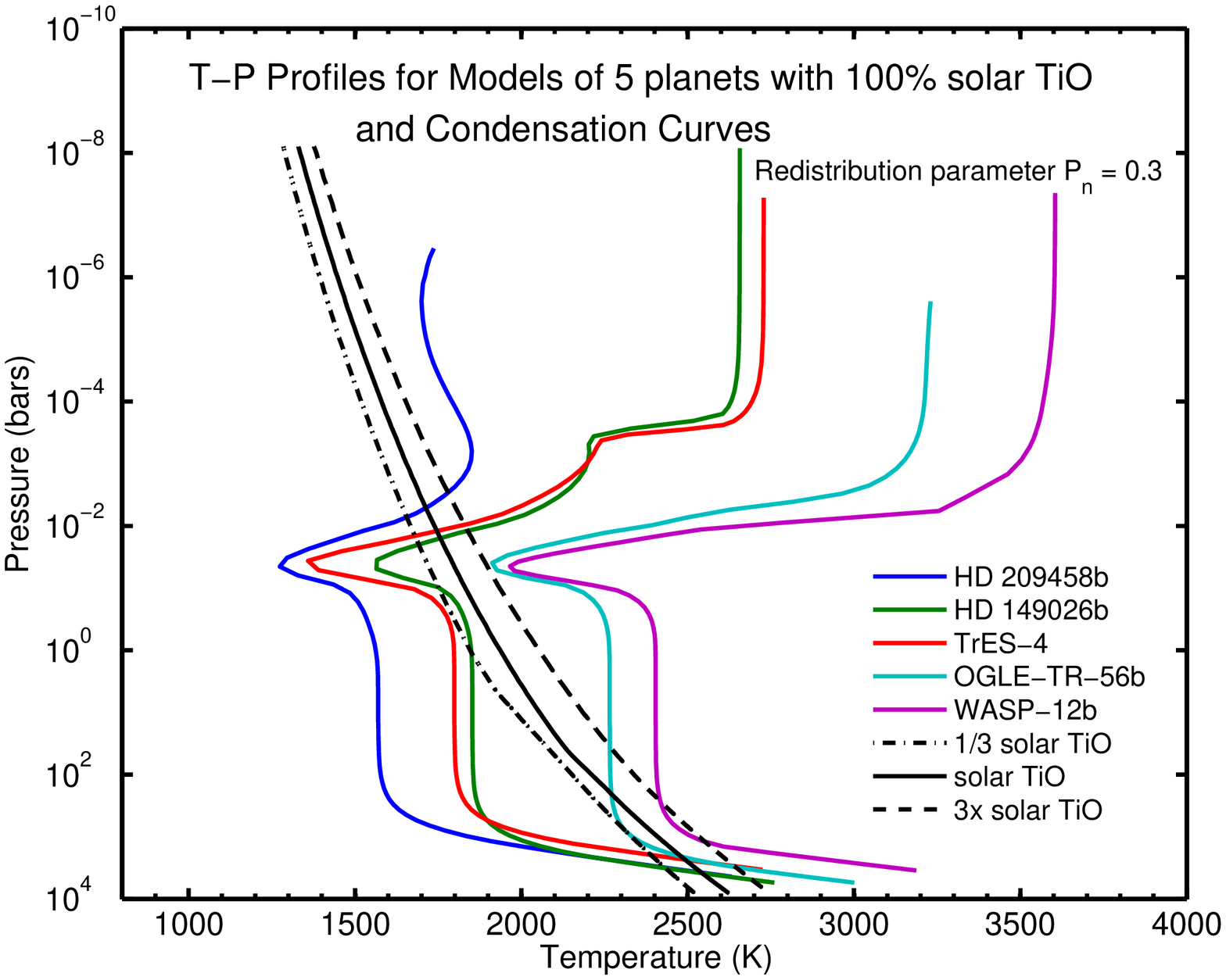}
\caption{Comparison of temperature-pressure profiles for models of
five planets with no TiO (left panel) and with solar abundance of TiO
(right panel).  This figure shows $T$-$P$ profiles for models without
(left panel) and with (right panel) the strong optical absorber TiO,
for HD209458b (blue), HD149026b(green), TrES-4 (red), OGLE-TR-56b
(cyan), and WASP-12b (magenta).  The redistribution parameter $P_n$ is
set equal to = 0.3.  Condensation curves (black) for titanium are
superposed, showing the locations of the condensation curves at 0.32
solar abundance (dashed-dotted line), solar abundance (solid line),
and 3.2 times solar abundance (dashed line).  The addition of TiO
heats the upper atmosphere and cools the lower atmosphere, because
more of the incident stellar flux is absorbed high in the atmosphere.}
\label{fig:5p_profiles}
\end{figure}
\end{landscape}

\begin{landscape}
\begin{figure}
\plottwo
{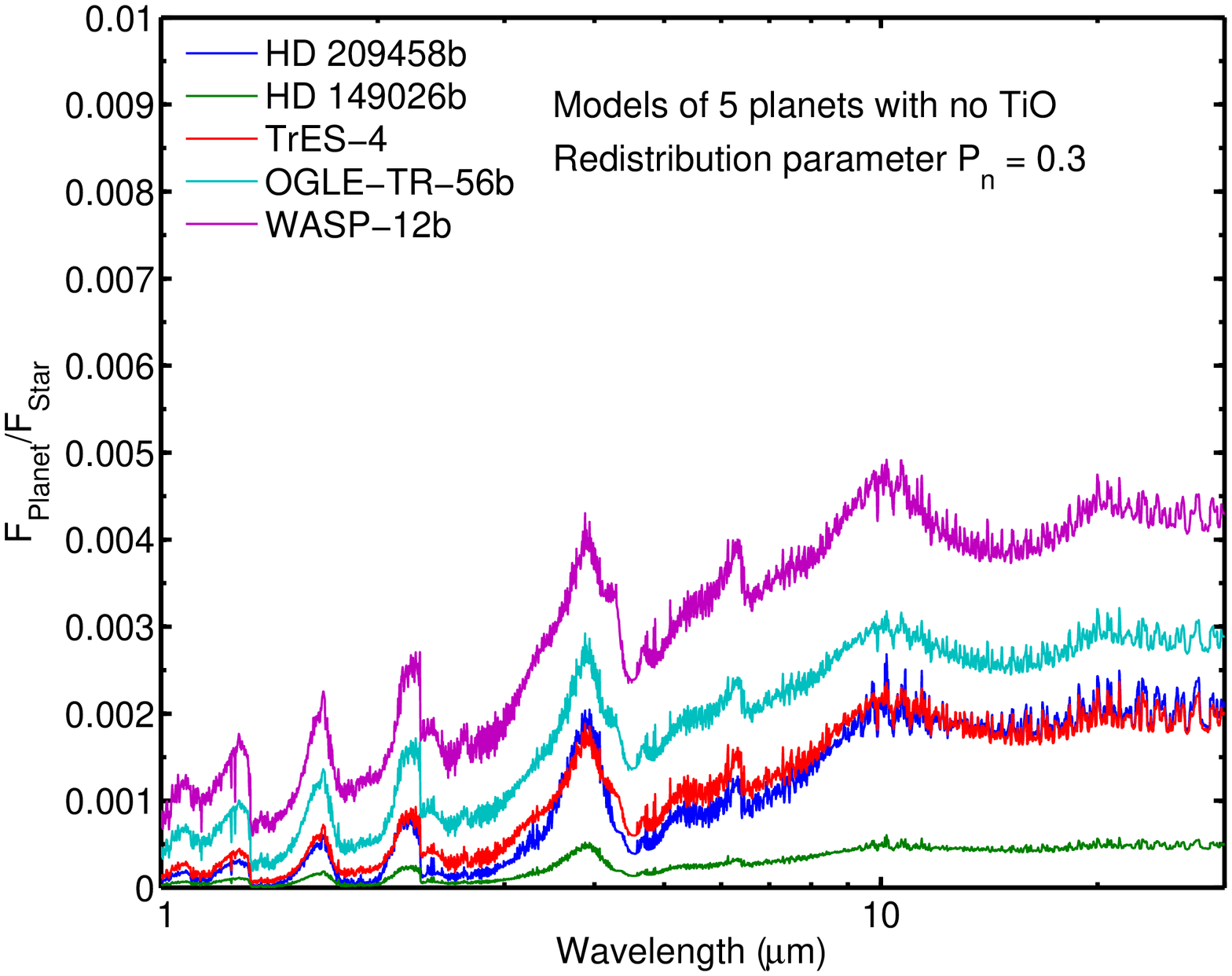}
{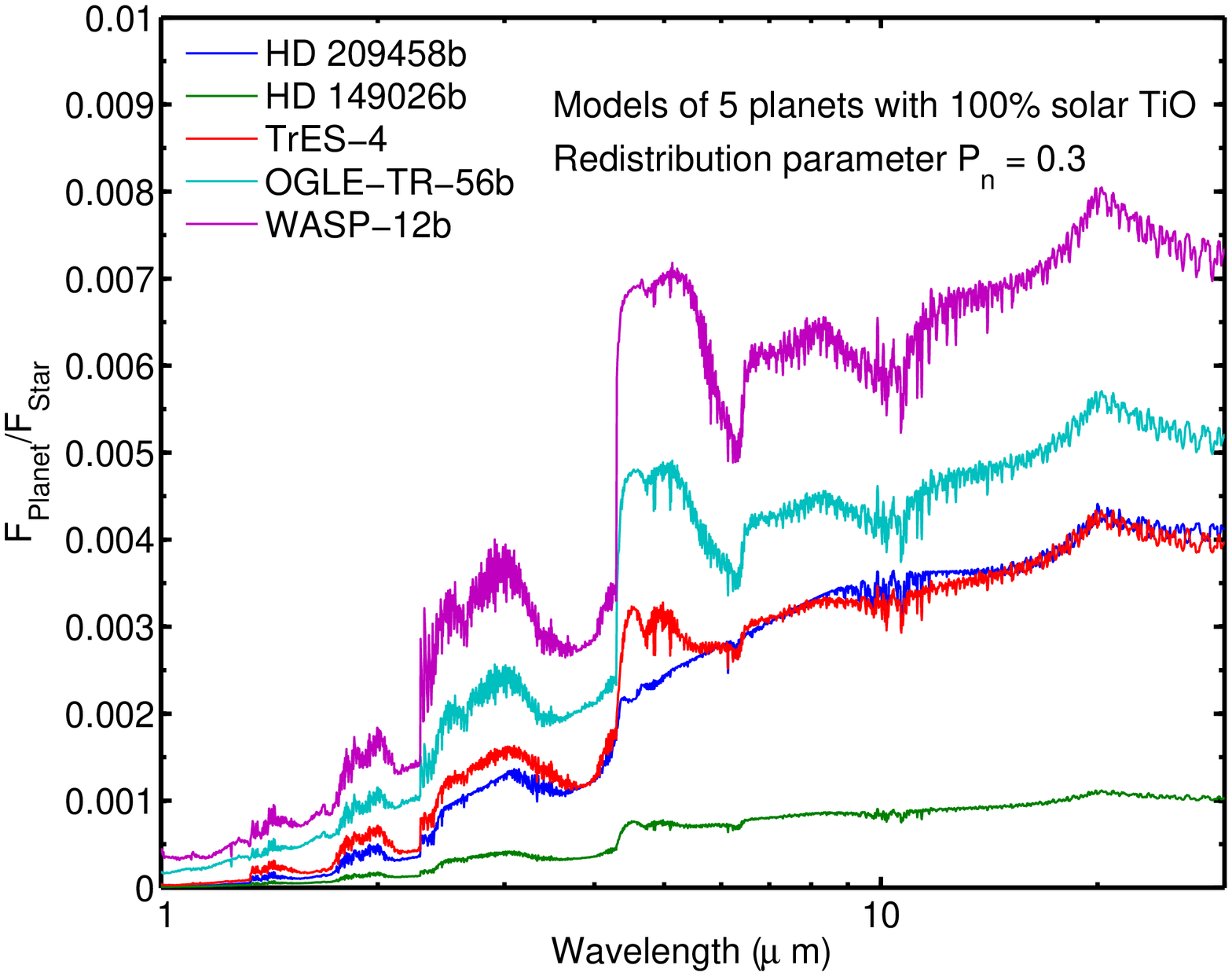}
\caption{Comparison of spectra for models of five planets with no TiO
(left panel) and with solar abundance of TiO (right panel).  This
figure presents the spectra that correspond to the $T$-$P$ profiles in
Fig.~\ref{fig:5p_profiles}, for the same five planets.  Here, adding
TiO increases the planet-star flux ratios in most of the IRAC range.
In particular, the flux ratios for IRAC 2 ($\sim$4.5~$\mu$m) and IRAC
3 ($\sim$5.8~$\mu$m) are significantly increased.}
\label{fig:5p_spectra}
\end{figure}
\end{landscape}

Titanium oxide, if present in a highly irradiated EGP's atmosphere,
can have a dramatic influence on both the temperature-pressure profile
and the emergent spectrum.  Figures~\ref{fig:HD209_spectra_TiO_kappa}
and \ref{fig:HD209_spectra_TiO} show this in the context of
HD~209458b.  However, there are other exoplanets where stellar
irradiation, and, hence, the influence of gaseous TiO, could be even
stronger.  This can be seen in the sharp contrasts displayed between
the two panels of both Figs.~\ref{fig:5p_profiles} and
\ref{fig:5p_spectra}.

Figures~\ref{fig:5p_profiles} and \ref{fig:5p_spectra} both show
models of five planets: 1) HD~209458b \citep{henry_et_al2000}, which
receives incident flux of $\sim 10^9 \rm~erg~cm^{-2}~s^{-1}$; 2)
HD~149026b \citep{sato_et_al2005}, which receives $\sim$2$\times 10^9
\rm~erg~cm^{-2}~s^{-1}$; 3) TrES-4 \citep{mandushev_et_al2007}, which
receives $\sim$2$\times 10^9 \rm~erg ~cm^{-2}~s^{-1}$; 4) OGLE-TR-56b
\citep{udalski_et_al2002d}, which receives $\sim$6$\times 10^9
\rm~erg~cm^{-2}~s^{-1}$; and 5) WASP-12b \citep{hebb_et_al2008}, which
receives $\sim$9$\times 10^9 \rm~erg~cm^{-2}~s^{-1}$ .  The mean
molecular weight $\mu$ is taken to be 2.3 for all planets, and
$\log_{10} g$ is taken to be 3.00, 3.19, 2.86, 3.27, and 3.04, for
HD~209458b, HD~149026b, TrES-4, OGLE-TR-56b, and WASP-12b,
respectively (where $g$ is in $\rm cm~s^{-2}$).  (See
Table~\ref{ta:Kzz_a} for a summary of these values.)

Figure~\ref{fig:5p_profiles} presents temperature-pressure profiles
for models of these planets with no TiO (left panel) and with solar
abundance of TiO (right panel).  Figure~\ref{fig:5p_spectra} portrays
spectra for the corresponding models.  We note that, as described in
\S\ref{ssec:num_meth}, models with TiO contain solar abundance of TiO
throughout the atmosphere, including in the cold trap; this procedure
should not produce large errors.  In the profile plots
(Fig.~\ref{fig:5p_profiles}) contain condensation curves for titanium
at 0.32 solar abundance (dashed-dotted black curve), solar abundance
(solid black curve), and 3.2 times solar abundance (dashed black
curve).  In the left panel, there is no cold-trap, because there is no
TiO.  In the right panel, as in Fig.~\ref{fig:cold_trap}, cold-trap
regions are found where each planet's profile is on the cold side of
the solar abundance condensation curve.  The solar abundance TiO
models for HD~209458b, HD~149026b, TrES-4, and OGLE-TR-56b all have
cold-trap regions.  Since the model for WASP-12b with solar abundance
TiO is never colder than the corresponding condensation curve, our
model predicts that WASP-12b would not have a day-side cold trap for
titanium if it is at or near solar abundance.
Figure~\ref{fig:5p_spectra} shows that in all five model planets,
solar abundance TiO produces a large thermal inversion in the upper
atmosphere (and cools the lower atmosphere).  This is reflected in the
spectrum as an increase in planet-star flux ratio throughout much of
the near infrared.

\begin{figure}
\plotone{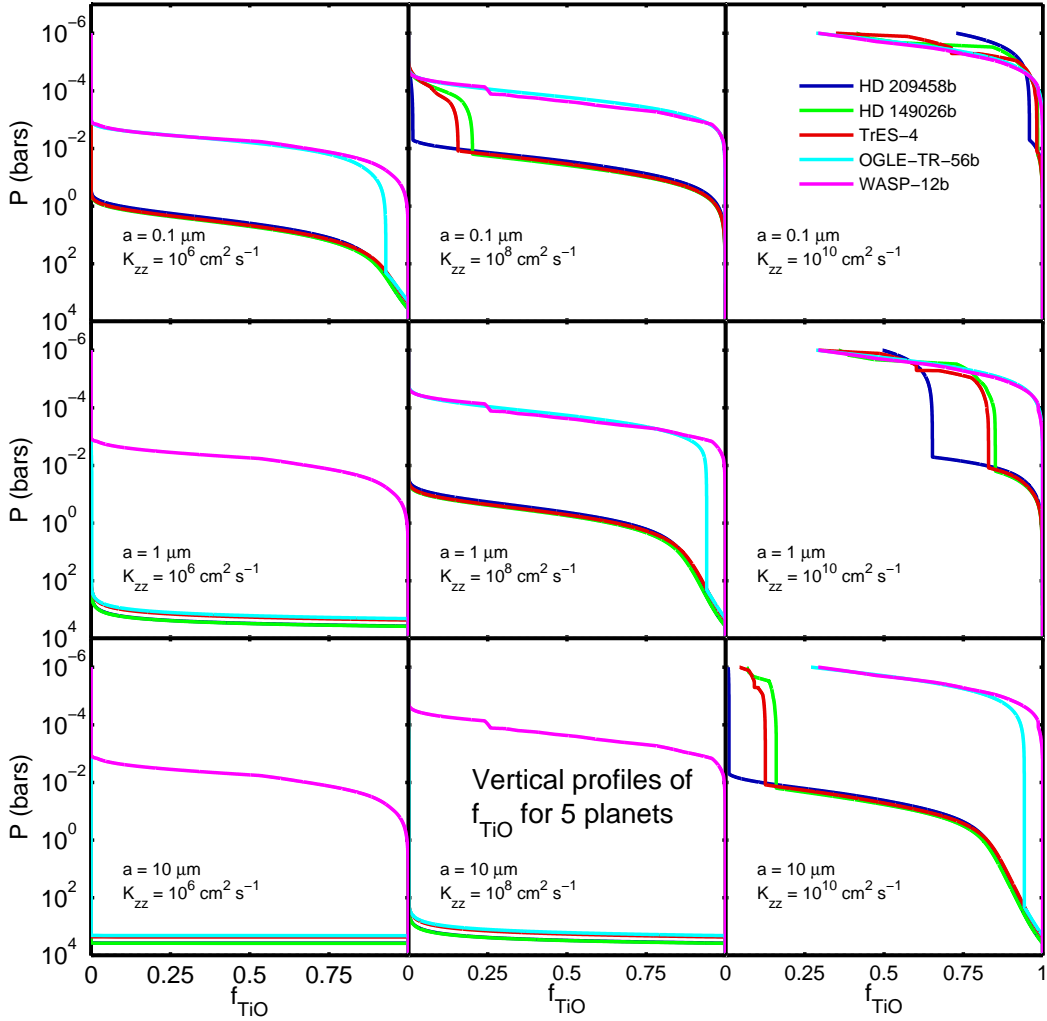}
\caption{Vertical profiles of $f_{\rm TiO}$ for five planets.  $f_{\rm
TiO}$ (abscissa) is shown as a function of pressure (ordinate) for
different combinations of particle size $a$ and turbulent diffusion
coefficient $K_{zz}$.  Left-to-right, $K_{zz}$ takes on the values
$10^6 \rm~cm^2~s^{-1}$, $10^8 \rm~cm^2~s^{-1}$, and $10^{10}
\rm~cm^2~s^{-1}$.  Top-to-bottom, $a$ varies from 0.1~$\mu$m to
1~$\mu$m to 10~$\mu$m.  HD~209458b, HD~149026b, TrES-4, OGLE-TR-56b,
and WASP-12b are represented, respectively, with dark blue, green,
red, cyan, and magenta (same color scheme as in
Figs.~\ref{fig:5p_profiles} and \ref{fig:5p_spectra}).  Since WASP-12b
has no day-side cold trap in our models, its curves are independent of
$a$.  At $K_{zz} = 10^6 \rm~cm^2~s^{-1}$, none of the model planets
has any appreciable amount of TiO in the upper atmosphere.}
\label{fig:f_of_P}
\end{figure}

\begin{landscape}
\begin{figure}
\vspace{-0.6in}
\plottwo
{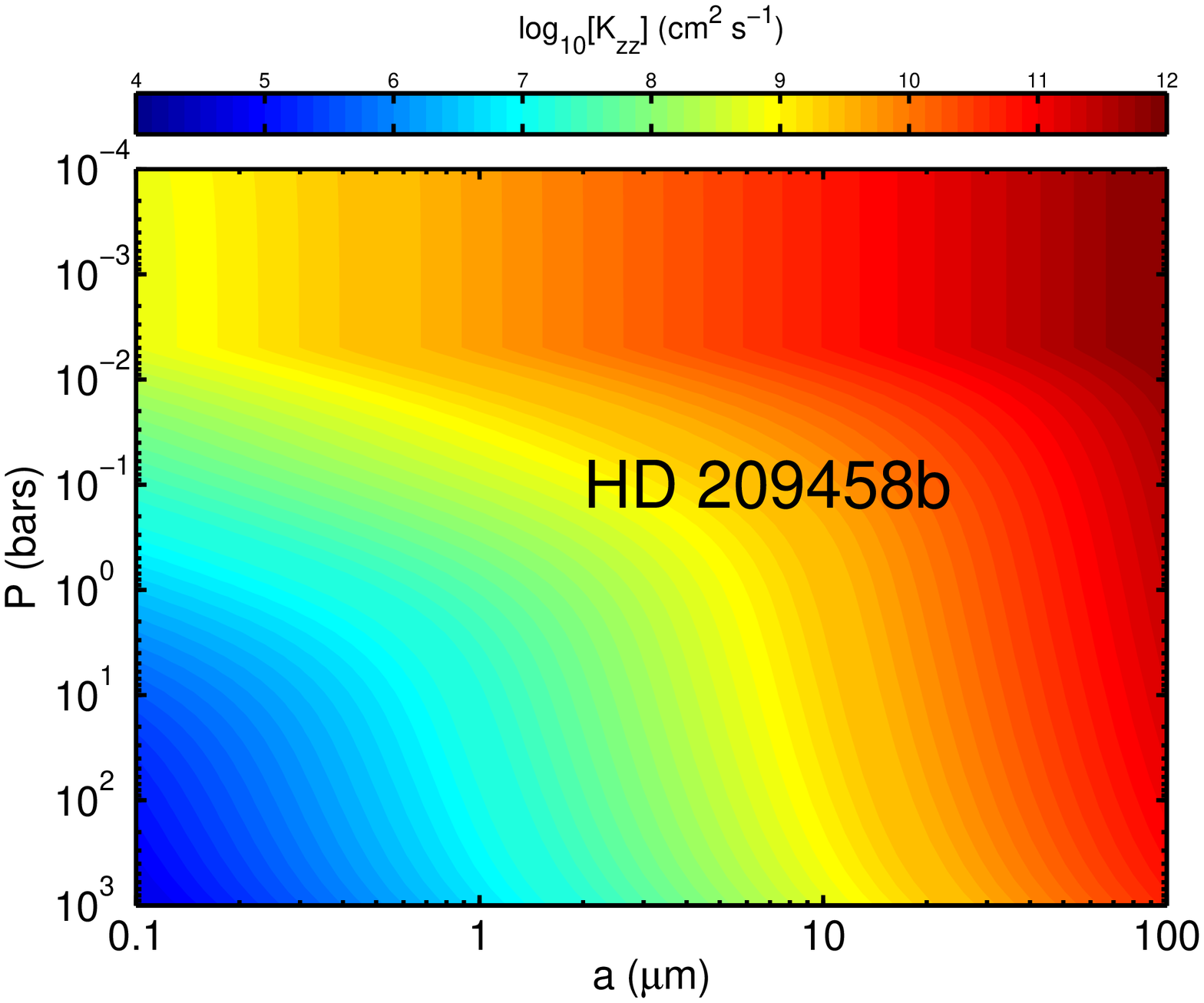}
{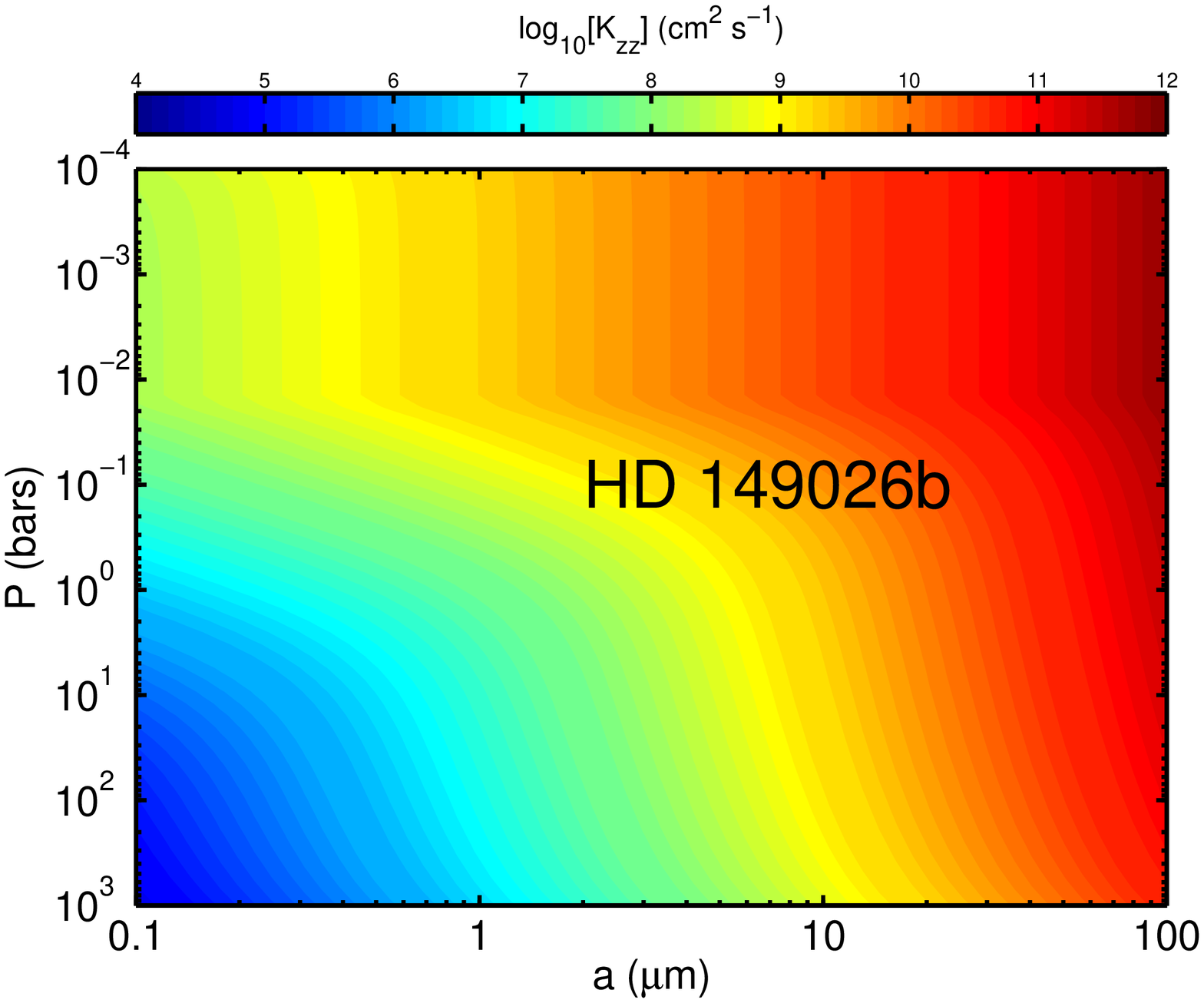}
\plottwo
{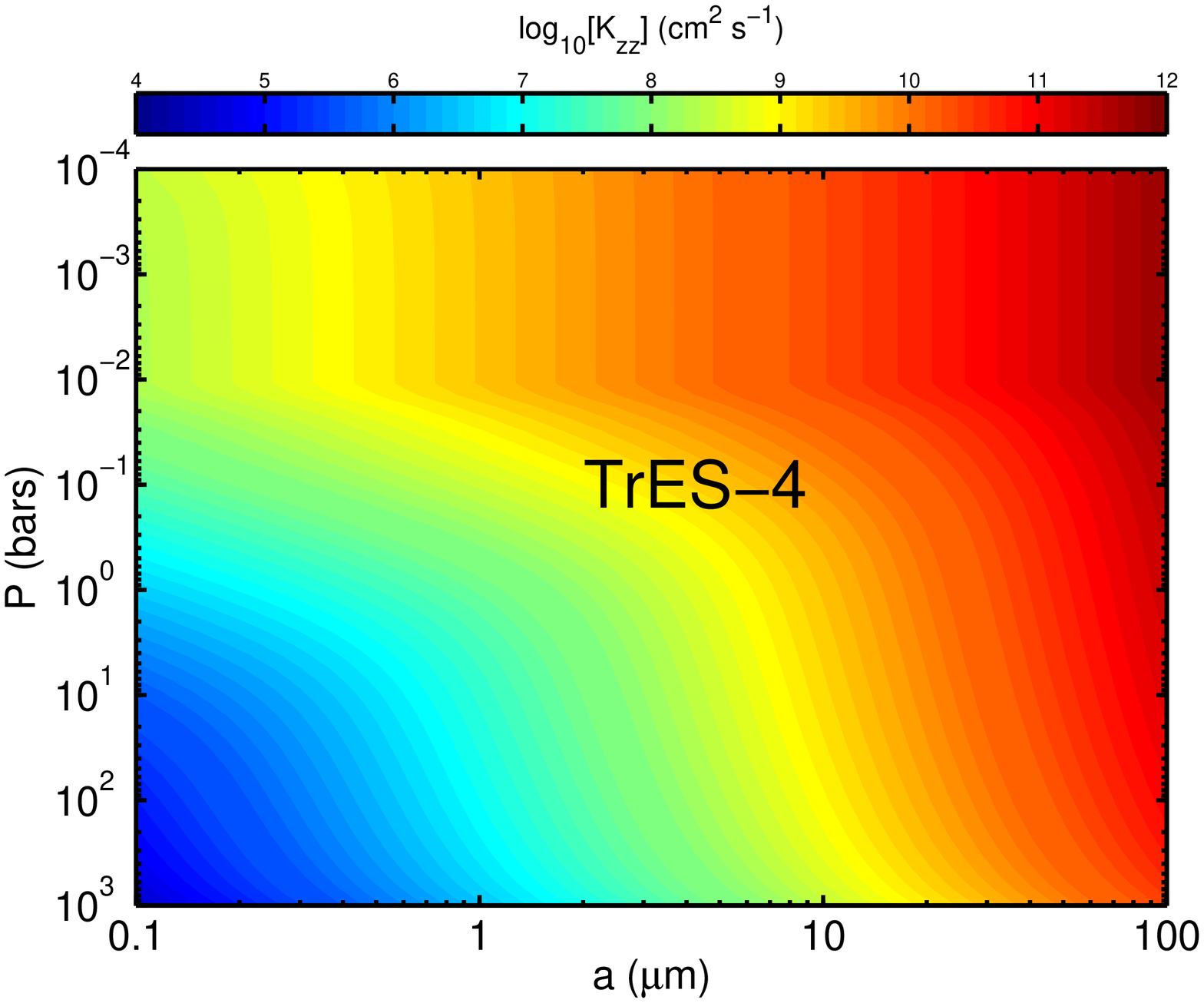}
{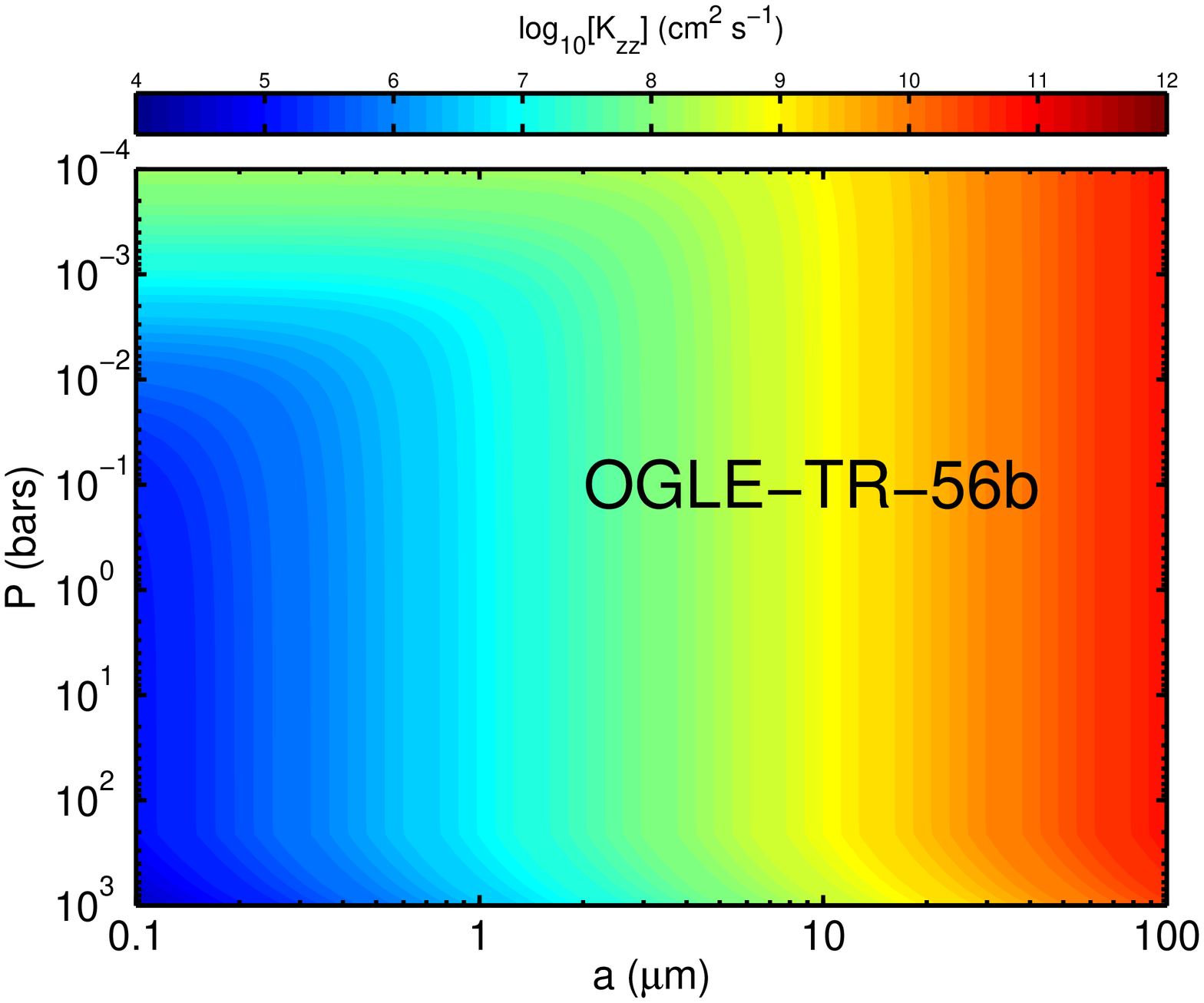}
\caption{Color-map of $\log[K_{zz}]$ (in $\rm cm^2~s^{-1}$) required
to achieve $f_{\rm TiO}=0.5$, as a function of $a$ and $P$ on the four
planets with day-side cold traps (HD~209458b: top-left; HD~149026b:
top-right; TrES-4: bottom-left; OGLE-TR-56b: bottom-right).}
\label{fig:Kzz_req_pretty}
\end{figure}
\end{landscape}

How much TiO is likely to survive the cold-traps on these planets and
to reach the upper atmospheres?  In Figs.~\ref{fig:f_of_P} through
\ref{fig:Kzz_req_pretty}, we address this question.

Figures \ref{fig:f_of_P}-\ref{fig:Kzz_req_pretty} all show the results
of integrating eq.~(\ref{eq:f2}) for different assumed values of $a$
and $K_{zz}$ (and for the different $T$-$P$ profiles of the different
planet models).  The $T$-$P$ profile (which is determined by finding a
radiative equilibrium solution to the radiative transfer equation in a
plane-parallel atmosphere, as described in \S\ref{ssec:num_meth} and
in the cited references) is related to altitude through the scale
height relationship of eq.~(\ref{eq:Hp}).  Terminal velocities $v_f$
are calculated with eqs.~(\ref{eq:vf}) and (\ref{eq:beta}).

Figure~\ref{fig:f_of_P} illustrates how much turbulent mixing is
required to achieve nonzero concentrations of TiO at the top of the
atmosphere of each of the five planets under consideration in this
paper.  This figure presents vertical profiles of $f_{\rm TiO}$ for
various combinations of $a$ (top-to-bottom: 0.1~$\mu$m to 1~$\mu$m to
10~$\mu$m) and $K_{zz}$ (left-to-right: $10^6 \rm~cm^2~s^{-1}$ to
$10^8 \rm~cm^2~s^{-1}$ to $10^{10} \rm~cm^2~s^{-1}$).  Since WASP-12b
has no day-side cold trap in our models, its curves are independent of
$a$.  To achieve nonzero concentrations of TiO at microbar pressures
requires values of $K_{zz}$ that are very high for a stably stratified
region such as the radiative part of an EGP's atmosphere ($K_{zz}
\gsim 10^{10} \rm~cm^2~s^{-1}$, even for WASP-12b, with no day-side
cold trap).  To achieve a nonzero concentration of TiO at millibar
pressures, the requirements on $K_{zz}$ are not quite so extreme, but
even still $K_{zz}$ must be $\gsim$10$^8 \rm~cm^2~s^{-1}$, even for
the smallest particle sizes, on all but the hottest planets
(OGLE-TR-56b and WASP-12b).  At $K_{zz} = 10^6 \rm~cm^2~s^{-1}$, none
of the planets has any appreciable amount of TiO in the upper
atmosphere, and only OGLE-TR-56b and WASP-12b have nonzero TiO above
$\sim$1~bar, even for $a=0.1~\mu$m.

As can be seen in
Figs.~\ref{fig:TiO_vs_VO}-\ref{fig:HD209_spectra_TiO} and
Fig.~\ref{fig:5p_profiles}, the thermal inversions reach maximum
temperatures at roughly millibar pressures.  This indicates that
$f_{\rm TiO}$ ought to be nonzero at these pressures.  But what
nonzero value is required?  Because, as Figs.~\ref{fig:f_at_mb} and
\ref{fig:f_at_mb_pretty} demonstrate, $f_{\rm TiO}$ transitions from
$\sim$0 to $\sim$1 in a fairly narrow range of $K_{zz}$, a reasonable
estimate is that, to cause a thermal inversion, $f_{\rm TiO}$ should
be $\sim$0.5 at $p\sim 10^{-3}$~bars.

Another way to visualize the analysis of Fig.~\ref{fig:f_of_P} is to
ask how $f_{\rm TiO}$ varies with $K_{zz}$ at a particular pressure
level.  Figure~\ref{fig:f_at_mb} portrays this relationship at a
millibar, on the same five planets, for condensate particle sizes
0.1~$\mu$m (top), 1~$\mu$m (middle), and 10~$\mu$m (bottom).  The
amount of turbulent mixing required to achieve a given $f_{\rm TiO}$
tends to increase with particle size.  The magenta curve for WASP-12b,
however, is independent of particle size and is, therefore, the same
in all three panels.  This figure shows that $f_{\rm TiO}$ is nearly 0
or nearly 1 for most values of $K_{zz}$, and makes the transition over
roughly an order of magnitude change in $K_{zz}$.  For condensates of
radius 10~$\mu$m, even OGLE-TR-56b requires $K_{zz} \sim 10^9
\rm~cm^2~s^{-1}$, while the other three planets with day-side cold
traps (HD~209458b, HD~149026b, and TrES-4) require $K_{zz}$ between
$10^{10}$ and $10^{11} \rm~cm^2~s^{-1}$.

Figure~\ref{fig:f_at_mb_pretty} displays how the millibar-level value
of $f_{\rm TiO}$ varies with $a$ and $K_{zz}$, for the four planets
with day-side cold traps.  The color contours indicate the mixing
ratio of TiO at $10^{-3}$~bars relative to the interior mixing ratio
of titanium.  The green band indicates the combinations required to
achieve the fiducial value of $f_{\rm TiO} \sim 0.5$.  This figure
shows that planets transition from $f_{\rm TiO} \sim 0$ to $f_{\rm
TiO} \sim 1$ in a fairly narrow range of values of $a$ and $K_{zz}$.
In general, smaller particles and more vigorous mixing produce larger
values of $f_{\rm TiO}$ at a millibar.

One more way to frame this result is to ask what value of $K_{zz}$ is
required to bring 50\% of the interior mixing ratio of TiO up to a
given pressure.  Figure~\ref{fig:Kzz_req_pretty} shows precisely this.
For the four planets with day-side cold traps, color contours indicate
the amount of turbulent mixing needed to achieve $f_{\rm TiO} = 0.5$,
as a function of $a$ and $P$.  At millibar pressures, each of these
planets requires from $\sim$10$^7 \rm~cm^2~s^{-1}$ of turbulent mixing
(in the case 0.1-$\mu$m particles in the cold trap of OGLE-TR-56b) to
$\sim$10$^{12} \rm~cm^2~s^{-1}$ (in the case of 100-$\mu$m particles
in the cold trap of HD~209458b).

Table~\ref{ta:Kzz_a} also presents the value of the eddy diffusion
coefficient $K_{zz}$ that is required to maintain $f_{\rm TiO} = 0.5$
above the day-side cold-trap, for various assumed condensate sizes.
0.1-$\mu$m particles require from $K_{zz}\sim 10^7 \rm~cm^2~s^{-1}$
(in the case of OGLE-TR-56b) to $\sim$10$^9 \rm~cm^2~s^{-1}$ (in the
case of HD~209458b), and these values increase roughly linearly with
particle radius.  1-$\mu$m particles require $K_{zz}\sim 10^7$ to
$\sim$10$^{10} \rm~cm^2~s^{-1}$; and 10-$\mu$m particles require
$K_{zz}\sim 10^9$ to $\sim$10$^{11} \rm~cm^2~s^{-1}$.

Since we lack constraints on both $a$ and $K_{zz}$, our uncertainty
spans many orders of magnitude.  Particle radii from 0.1~$\mu$m to
30~$\mu$m or more are not implausible, nor are values of $K_{zz}$ from
10$^2$ to 10$^9 \rm~cm^2~s^{-1}$, possibly even greater.  Only for a
narrow range of $a$-$K_{zz}$ space is the upper atmosphere abundance
of TiO at all sensitive to $a$ and $K_{zz}$.  For most of parameter
space, there is either not nearly enough upper atmosphere TiO to cause
thermal inversions, or easily enough.

\section{Caveats}
\label{sec:caveats}
What should we conclude from the results in \S\ref{sec:results}?  In
this section, we address a few additional complications of our
analysis.  First, in \S\ref{ssec:singlezone}, we consider how a more
sophisticated analysis would treat the single-zone model, which
considers motions only in the vertical direction, presented in
\S\ref{sec:model}.  Then, in \S\ref{ssec:coupling}, we qualitatively
describe how horizontal winds, which effectively couple the day side
of a planet to colder parts (including both the night side and the
polar regions), influence the day-side upper atmosphere abundance of
TiO.

\subsection{Caveats for the Single-Zone Model}
\label{ssec:singlezone}
First, the models presented in \S\ref{sec:results} assume either solar
abundance of TiO or no TiO, but a planet's interior mixing ratio of
titanium might be super-solar.  If so, then $f_{\rm TiO}$ could be
lower than 0.5 while still maintaining a 50\% solar mixing ratio of
titanium in the upper atmosphere.  If a planet's interior titanium
abundance were twice solar, for instance, $f_{\rm TiO}$ could be 0.25
for a 50\% solar mixing ratio of titanium at the top of the cold trap.
Nonetheless, the band in $a$-$K_{zz}$ space in which $f$ changes by a
factor of two from, say, 0.5 to 0.25, is fairly narrow in comparison
to the range of plausible values.  Even an order of magnitude change
from 0.5 to 0.05 corresponds to a fairly modest change in $K_{zz}$ of
a factor of $\sim$3.

It is also conceivable that we dismiss too quickly the possibility
that VO contributes significantly to thermal inversions.  According to
\citet{sharp+burrows2007}, vanadium condenses at somewhat lower
temperature than titanium.  All else being equal, its cold-trap would,
therefore, be smaller.  Still, since even 10 times solar abundance of
VO does not produce as large a thermal inversion as has been inferred
from observations of HD~209458b it is unlikely that VO could play a
key role in producing thermal inversions.

There are two ways in which our analysis is not self-consistent, both
related to the false assumption in most of
Figs.~\ref{fig:TiO_vs_VO}-\ref{fig:5p_profiles} that TiO is present at
constant mixing ratio throughout a planet's entire atmosphere.  First,
we use the temperature-pressure profiles that result from this
assumption to find where the cold-traps are.  Second, we also use
condensation curves calculated based on this assumption in finding the
cold-traps.  In reality, as (the non-self-consistently generated)
Fig.~\ref{fig:f_of_P} demonstrates, the mixing ratio of TiO decreases
with altitude.  Lower abundance of titanium condenses at lower
temperature for a given pressure.  As a result, the condensation
curves of a self-consistent analysis would have shallower slopes that
reflect the decrease in mixing ratio of TiO at lower pressures.  A
more sophisticated model would self-consistently take into account the
progressive depletion of TiO in calculating both $T$-$P$ profiles and
condensation curves.

Finally, one might ask whether our use of day-side average models
could mask the presence of local conditions near the substellar point
that are hot enough not to have cold traps.  For the four planets that
our model predicts have cold traps (HD 209458b, HD 149026b, TrES-4,
and OGLE-TR-56b), could the intense insolation near the substellar
points raise the temperature enough that the cold traps don't exist at
these local hot spots?  As described in \citet{burrows_et_al2008b}
Appendix D, the average conditions correspond to the ring around the
substellar point with direction cosine $\mu=2/3$.  However, local
conditions at the substellar point (where irradiation is 50\% greater
than the average used) are such that on HD 209458b, HD 149026b, and
TrES-4, the substellar point flux still receives less flux than the
day-side average flux received by OGLE-TR-56b, a planet that according
to our model has a cold-trap.  This suggests that even local 2D models
of these 3 planets would have no part of the day side hot enough to
avoid having a cold trap.  Moreover, parts of the atmosphere with
$\mu<2/3$ would have even larger cold traps than predicted by our
models.  Even if some planets, such as perhaps OGLE-TR-56b, do not
have $T$-$P$ profiles corresponding to cold trap conditions at their
substellar points, planetary winds would prevent gaseous TiO from
remaining at such hot locations for long.  Coupling between parts of a
planet with different local conditions is the subject of
\S\ref{ssec:coupling}.

\subsection{Other Depletion Regions}
\label{ssec:coupling}
The actual process by which planetary atmospheres achieve their
spatially and temporally dependent chemical distributions is more
complicated than the model presented above.  In addition to molecular
and turbulent diffusion, planetary-scale winds advect gas and
condensates from the day side to the night side and
back\footnote{\citealt{showman_et_al2008b} suggest that the vertical
wind speeds (30$\rm~m~s^{-1}$) in their simulations might be
sufficient to prevent significant night-side settling of condensates
smaller than 30$\rm~\mu m$ on each circulation.  It is possible,
however, that night-side condensate settling over a $\sim$ billion
year timescale drains the upper atmosphere of TiO.}, equator to pole
and back, and across different altitudes.  The chemical composition of
advected Lagrangian parcels of air can change in response to the local
temperature and pressure conditions that they experience in the course
of their circulation.  The TiO content above the day-side cold-trap,
therefore, depends on a variety of timescales: the timescale for zonal
circulation, $\tau_{\rm circ} \sim \pi R_p/u$, where $u$ is the zonal
wind speed; the timescale for gravitational settling of condensates,
$\tau_{\rm settle} \sim \Delta z/\beta {v_f}_{\rm Stokes}$; the
timescale for chemical transitions -- the formation and destruction of
condensates, $\tau_{\rm chem}$; and the timescale for heating and
radiatively cooling, $\tau_{\rm rad}$.

Consider the influence of the night-side cold-trap.  The basic
formalism for the vertical distribution of a species on the night side
(in the absence of advection) is the same as that on the day side.
Because the night side has its own temperature-pressure profile (that
is on the condensed side of the condensation curve of Ti for most of
the atmosphere), it will have its own (advection-free) vertical
distribution of species $i$, here labeled $n_{i,\rm n}$.  Coupling
between the day and night sides may be described quantitatively by
adding a source/sink term to eq.~(\ref{eq:diffusion_eq}).  This term
ought to be proportional to the difference between the number density
on the night and day sides, and inversely proportional to the
circulation time.  In the context of a simple model, suppose that each
$z$-level on the day side is carried by circulatory winds to one
unique level on the night side (and vice versa).  Let $Z_{\rm n}(z)$
be the monotonic, invertible function that maps day-side altitude $z$
to the night-side altitude to which it is coupled.  The revised
version of eq.~(\ref{eq:diffusion_eq}), then, is the following:
\begin{eqnarray}
\nonumber \frac{\partial n_i[z]}{\partial t} & = & \frac{\partial}{\partial z} \left\{ D_i \left( \frac{\partial n_i}{\partial z} + \left( \frac{\partial \ln T}{\partial z} + \frac{m_i g}{kT} \right)n_i \right) \right. \\
\label{eq:diffusion_eq_coupling} & & \left. + K_{zz} \left( \frac{\partial n_i}{\partial z} + \left( \frac{\partial \ln T}{\partial z} + \frac{\mu m_p g}{kT} \right) n_i \right) \right\}  + \frac{n_{i,\rm n} \left[ Z_{\rm n}(z)\right] - n_i[z]}{\tau_{\rm circ}}.
\end{eqnarray}
The corresponding equation for the night side looks nearly identical,
except 1) a night-to-day coupling function, which is the inverse
function of the day-to-night function ($Z_{\rm d} \equiv {Z_{\rm
n}}^{-1}$), takes the place of $Z_{\rm n}$; and 2) the sign of the
$1/\tau_{\rm circ}$ term is reversed.

For the reasons discussed in \S\ref{ssec:model}, we still may assume
$\partial n_i/\partial t = 0$.  Now, a vertical integration yields
\begin{eqnarray}
\nonumber D_i \left( \frac{\partial n_i}{\partial z} + \left( \frac{\partial \ln T}{\partial z} + \frac{m_i g}{kT} \right)n_i \right) + K_{zz} \left( \frac{\partial n_i}{\partial z} + \left( \frac{\partial \ln T}{\partial z} + \frac{\mu m_p g}{kT} \right) n_i \right) & &  \\
\label{eq:diffusion_eq_coupling2} + \frac{1}{\tau_{\rm circ}}\int_0^z \left(n_{i,\rm n}[Z_{\rm n}(\zeta)] - n_i[\zeta] \right)d\zeta & =  & 0 \, .
\end{eqnarray}

We briefly consider the qualitative properties of this equation.  An
important unknown is the nature of the coupling function $Z_{\rm n}$.
This function depends on the trajectories followed by Lagrangian fluid
elements that start at different altitudes in the course of their
circulation around a planet.  Although neither entropy nor pressure is
likely to be strictly conserved, are flow patterns best described as
isentropic or isobaric?

Since a highly irradiated EGP's night side lacks the day side's
intense external irradiation, its static stability is less and its
radiative-convective boundary extends up to lower pressures than the
corresponding boundary on the day side.  Furthermore, the cooler night
side has a smaller pressure-scale height.  For both of these reasons,
if the circulation is nearly isobaric, $n_{i,\rm n}[Z_{\rm n}(z)]$
could actually be greater than $n_{i,\rm n}[z]$ for $z$ in the
day-side cold-trap region.  The night-side cold-trap could, therefore,
actually serve as a {\it source} for the day side, and might make it
easier for TiO to reach the upper atmosphere.

In contrast, if circulation nearly follows isentropes, parcels on the
day side will travel to locations on the night side that are a greater
altitude above the convective zone than they had been on the day side
(i.e., $Z_{\rm n}[z] > z$).  In this case, the night side is purely a
sink.

\citet{showman_et_al2008} define the radiative timescale as the
$e$-folding time for temperature perturbations (of magnitude $\delta
T$) of a $T$-$P$ profile to decay: $\tau_{\rm rad} \sim \delta T {\rho
c_p}/({dF/dz})$, where $\rho$ is mass density, $c_p$ is the specific
heat at constant pressure, and $F$ is the net vertical flux.  Note
that in radiative equilibrium $dF/dz$ equals zero.  Therefore, the
radiative timescale as defined above makes sense only in the case of a
perturbed atmosphere.  In regimes of $T$-$P$ space where $\tau_{\rm
rad}$ is large relative to $\tau_{\rm circ}$, a fluid parcel will not
change its entropy by much during the course of its circulation.  For
HD 209458b, equatorial winds of $\sim$1$\rm~km~s^{-1}$ imply a
circulation timescale of $\tau_{\rm circ} \sim 3\times 10^5~\rm s$.
According to the analysis of \citet{showman_et_al2008}, on this planet
$\tau_{\rm rad} > \tau_{\rm circ}$ at pressures deeper than 1~bar.  In
the deep regions of the day-side cold-trap, therefore, the night side
is likely to be sink.  If winds travel at speeds close to the local
speed of sound in the deep atmosphere, shocks might make the
circulation non-isentropic, despite the long radiative timescale,
although we note that circulation models generally predict wind speeds
well below the sound speed at pressures of 1 bar and greater
\citep{menou+rauscher2008,showman_et_al2008b}.

The other influence of coupling to the night side is that parcels of
air from the upper atmosphere of the day side, above the cold-trap
region, will be advected to cooler regions on the night side.  If
condensates form nearly instantaneously ($\tau_{\rm chem} \ll
\tau_{\rm circ}$), then any TiO above the day side's cold-trap
condenses and begins to settle while on the night side.  On HD
209458b, the terminal velocity of condensates is $\sim$4${\rm
~cm~s^{-1}}\beta(a/10{~\mu \rm m})^2$.  Condensates of radius $10
\rm~\mu m$, therefore, fall $\Delta z_{\rm settle} = \tau_{\rm circ}
\beta {v_f}_{\rm Stokes} \sim 10\beta \rm~km$ while on the night side.

This settling process is countered by turbulent diffusion.  The
characteristic distance that the condensates are lofted by turbulent
diffusion in a circulation time is $\Delta z_{\rm turb} \sim
\sqrt{\tau_{\rm circ} K_{zz}}$.  If $\Delta z_{\rm turb} < \Delta
z_{\rm settle}$, the day side's upper atmosphere is steadily depleted
of TiO, as it condenses on the night side and settles into the
cold-trap region.  If $\Delta z_{\rm turb} > \Delta z_{\rm settle}$,
this process might not significantly alter the upper atmosphere TiO
mixing ratio.

\section{Summary and Conclusions}
\label{sec:conc}
\begin{figure}
\plotone{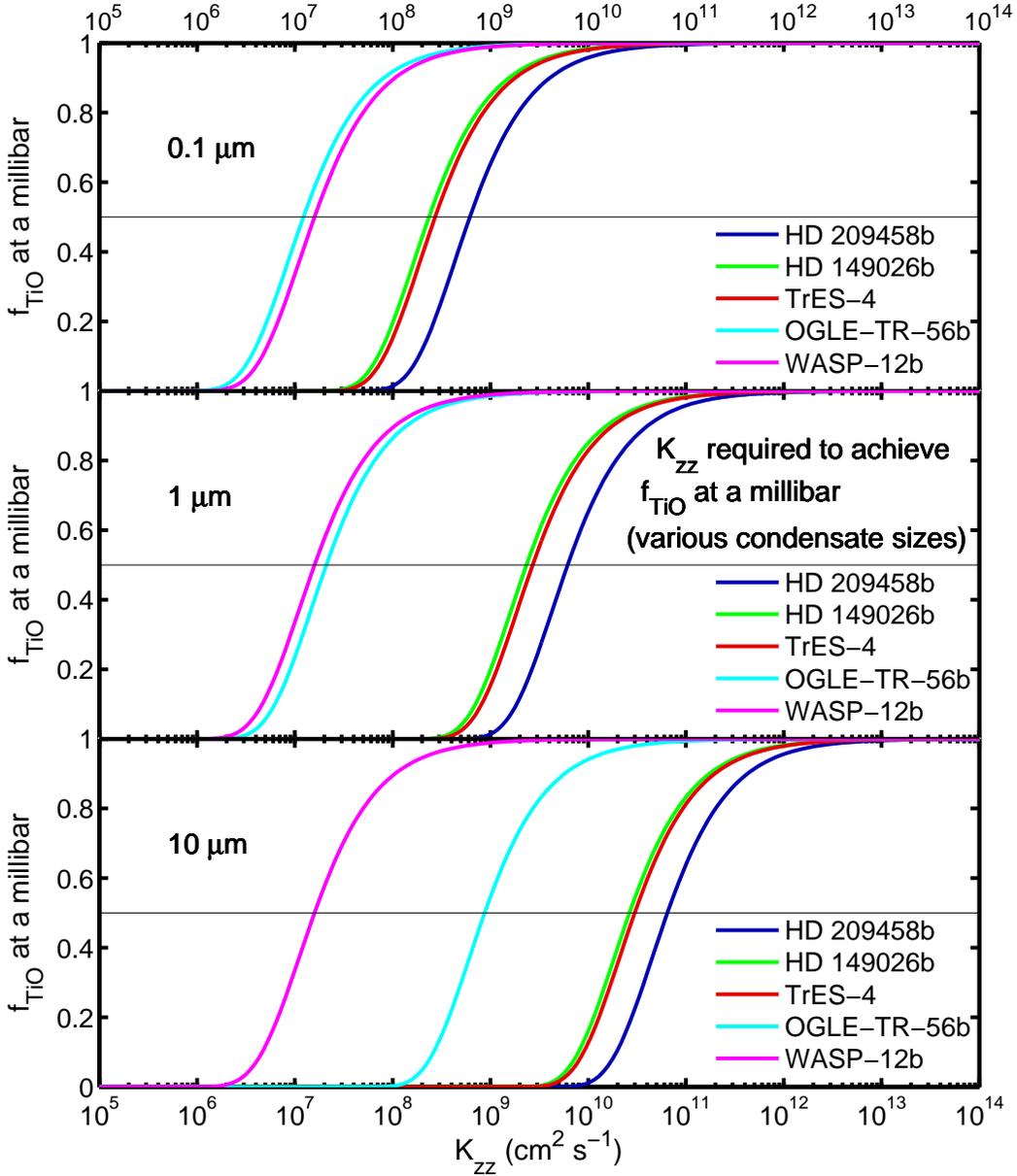}
\caption{$K_{zz}$ required to achieve $f_{\rm TiO}$ at $p = 10^{-3}$
bars, for five planets, for various condensate particle sizes.  For
condensates ranging from 0.1~$\mu$m (top panel) to 10~$\mu$m (bottom
panel), $f_{\rm TiO}$ is plotted as a function of $K_{zz}$ on each of
the five planets considered in this paper (same color scheme as in
previous figures).  The magenta curve for WASP-12b is independent of
particle size and so is identical in all three panels.  A horizontal
black line is shown in each plot at $f_{\rm TiO} = 0.5$ to aid the eye
in identifying the value of $K_{zz}$ that is required to achieve this
fiducial relative mixing ratio. }
\label{fig:f_at_mb}
\end{figure}

As has been published, an additional upper atmosphere absorber in the
optical can produce the thermal inversions inferred from observations.
There is an oft-quoted hypothesis in the literature that the strong
optical absorbers TiO and VO are responsible for these thermal
inversions.  Here, we have studied the viability of this hypothesis,
with a radiative-convective radiative-transfer model and a model of
particle settling in the presence of turbulent and molecular
diffusion.  We applied these models to five highly irradiated EGPs:
HD~209458b, HD~149026b, TrES-4, OGLE-TR-56b, and WASP-12b,
parameterizing our results (see Table~\ref{ta:Kzz_a}) in terms of
sizes of condensed particles in cold-trap regions and the strength of
eddy diffusion.  Our most important findings are the following:
\begin{itemize}
\item
It is unlikely that VO plays a role in producing an upper atmosphere
thermal inversion.
\item
In four of the five planets considered, a TiO cold-trap is likely to
exist between the hot convection zone and the hot upper atmosphere on
the irradiated day sides of the planets.  The titanium that is present
in such cold-traps is likely to be sequestered in a variety of
condensates that settle much more strongly than does gaseous TiO.  The
only planet that does not have a day-side cold-trap is WASP-12b, which
receives at least 50\% more irradiation than any other known planet.
\item
Macroscopic mixing is essential to the TiO hypothesis.  Without
macroscopic mixing processes, such as turbulent diffusion, a heavy
molecular species such as TiO will not be present in a planet's upper
atmosphere.  Although WASP-12b, for instance, has no cold-trap in our
analysis, it still requires turbulent mixing of $\sim$10$^7
\rm~cm^2~s^{-1}$ (see Fig.~\ref{fig:f_at_mb}) if TiO is to be abundant
above a millibar in its upper atmosphere.
\item
Planetary-scale winds that couple the day side of a planet to the
colder night side are likely to make it even more difficult than is
indicated by the models in this paper for enough titanium to reach the
upper atmosphere for TiO to produce a thermal inversion.
\end{itemize}

Finally, we estimate how much macroscopic mixing is required to loft
enough condensed titanium above the day-side cold-trap for TiO to
cause a significant inversion.  If titanium is sequestered in
condensates of radius $a$, then our model predicts that, for gaseous
TiO to be present in the upper atmosphere at sufficient quantity to
cause thermal inversion, $K_{zz}$ must have the following values on
the following planets: 1) on HD~209458b, $K_{zz}$ must be
$\gsim$6.2$\times 10^8 (a/1{\rm{~\mu}m}) \rm ~cm^2~s^{-1}$; 2) on
HD~149026b, $K_{zz}$ must be $\gsim$2.4$\times 10^8 (a/1{\rm{~\mu}m})
\rm ~cm^2~s^{-1}$; 3) on TrES-4, $K_{zz}$ must be $\gsim$2.7$\times
10^8 (a/1{\rm{~\mu}m}) \rm ~cm^2~s^{-1}$; 4) on OGLE-TR-56b, $K_{zz}$
must be $\gsim$1.2$\times 10^7 \rm ~cm^2~s^{-1}$ for $a = 0.1~\mu$m,
$\gsim$2.1$\times 10^7 \rm ~cm^2~s^{-1}$ for $a = 1~\mu$m, and
$\gsim$8.7$\times 10^7 \rm ~cm^2~s^{-1}$ for $a = 10~\mu$m; and 5) on
WASP-12b, $K_{zz}$ must be $\gsim$1.6$\times 10^7 \rm ~cm^2~s^{-1}$.
The analysis that leads to these estimates neglects the effect of the
night-side cold-trap, and, therefore, these values should be taken as
lower limits.  Because both $K_{zz}$ and $a$ are currently unknown, it
remains to be seen whether TiO can indeed be responsible for thermal
inversions in highly irradiated EGPs.  Though our results suggest that
the TiO hypothesis might be problematic, they provide a framework in
which to assess it, given improved estimates of $K_{zz}$ and $a$ in
the future.

\acknowledgments
We thank Laurent Ibgui, Ivan Hubeny, Jason Nordhaus, Jonathan
Mitchell, and Bruce Draine for useful discussions.  We acknowledge our
anonymous referee for numerous helpful comments.  This study was
supported in part by NASA grant NNX07AG80G.  We also acknowledge
support through JPL/Spitzer Agreements 1328092, 1348668, and 1312647.


\bibliography{biblio.bib}

\newpage

\begin{deluxetable}{l||rr|crrr}
\tablecaption{Planetary gravitational acceleration, stellar irradiation, and
required $K_{zz}$}
\tablehead{\colhead{Planet} & \colhead{$g$} & \colhead{$F_*$} & \colhead{} & \colhead{} & \colhead{required $K_{zz}$} & \colhead{}\\
\colhead{} & \colhead{$\rm cm~s^{-2}$} & \colhead{$\rm erg~cm^2~s^{-1}$} & \colhead{$a =$} & \colhead{$0.1 \rm~{\mu}m$} & \colhead{$1 \rm~{\mu}m$} & \colhead{$10 \rm~{\mu}m$}}
\startdata
HD~209458b  & 1000 & 1.0 & & $6.2\times 10^8$  & $6.2\times 10^9$  & $6.5\times 10^{10}$ \\
HD~149026b  & 1560 & 2.2 & & $2.4\times 10^8$  & $2.3\times 10^9$  & $2.6\times 10^{10}$ \\
TrES-4      &  721 & 2.4 & & $2.7\times 10^8$  & $2.7\times 10^9$  & $3.0\times 10^{10}$ \\
OGLE-TR-56b & 1850 & 5.5 & & $1.2\times 10^7$  & $2.1\times 10^7$  & $8.7\times 10^8$ \\
WASP-12b    & 1090 & 9.3 & & $^*1.6\times 10^7$  & $^*1.6\times 10^7$  & $^*1.6\times 10^7$ \\
\enddata
\label{ta:Kzz_a}
\tablecomments{ This table gives planetary $g$ and stellar flux
($F_*$), and values of $K_{zz}$ (in $\rm cm^2~s^{-1}$) required to
achieve $f_{\rm TiO} = 0.5$ above the cold trap, for particles sizes
of 0.1~$\mu$m to 10~$\mu$m.  The asterisks in the last row are because
WASP-12b has no day-side cold trap.  The required value of $K_{zz}$,
therefore, is independent of condensate particle size.}
\end{deluxetable}

\newpage

\newpage
\clearpage
\clearpage
\newpage
\newpage
\clearpage
\newpage
\newpage
\clearpage
\newpage
\newpage
\begin{landscape}
\vspace{-0.3in}
\begin{figure}
\vspace{-0.7in}
\plottwo
{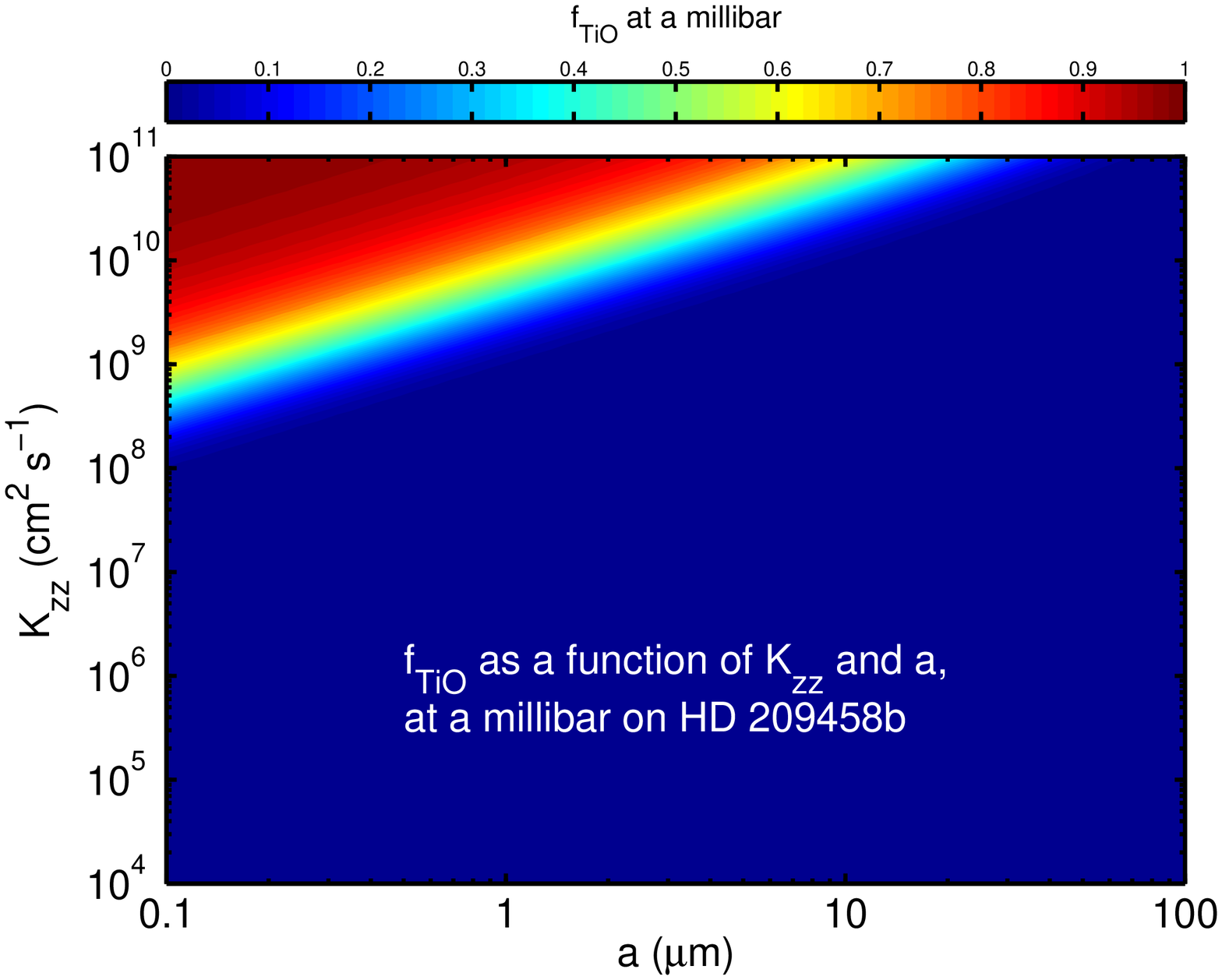}
{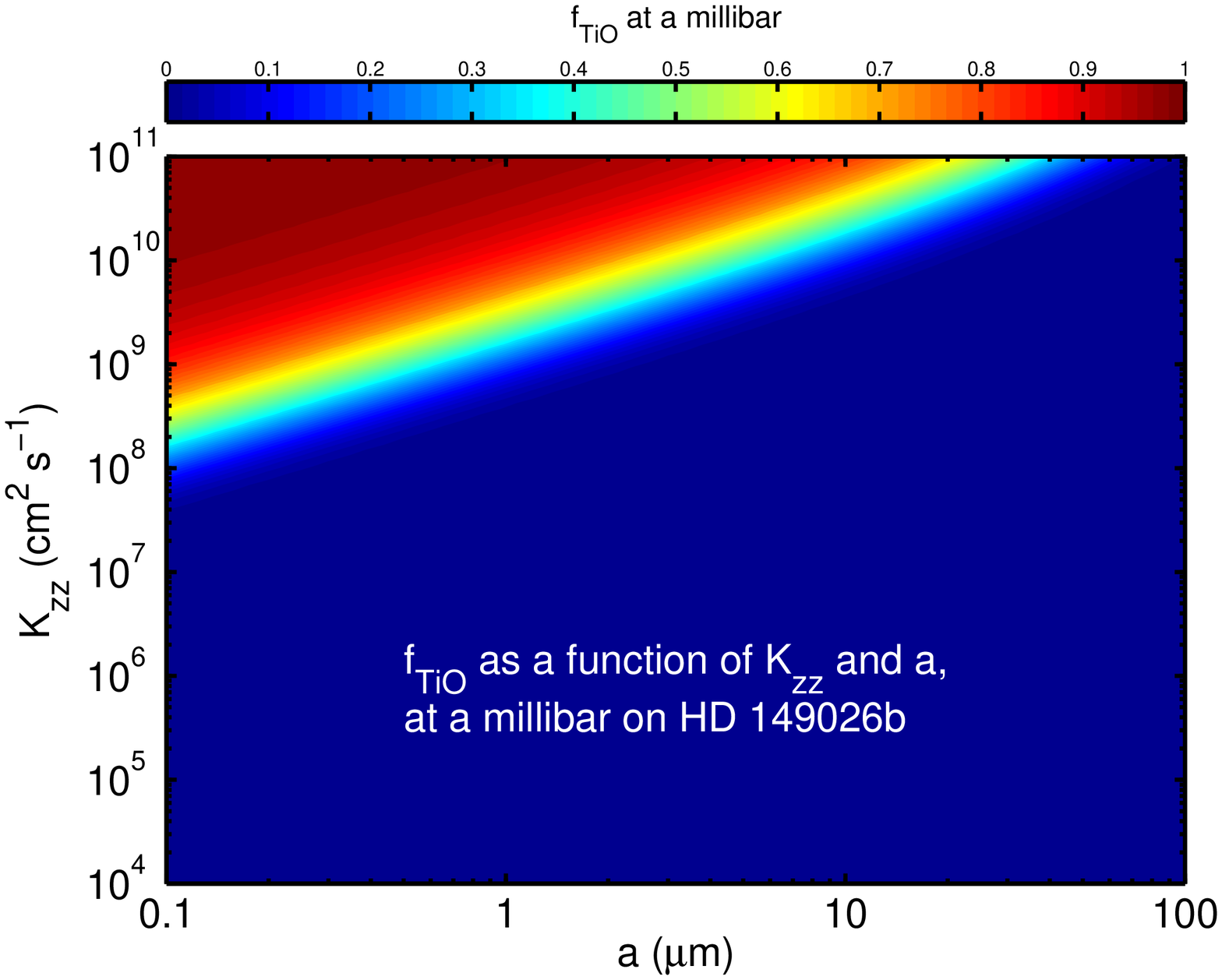}
\plottwo
{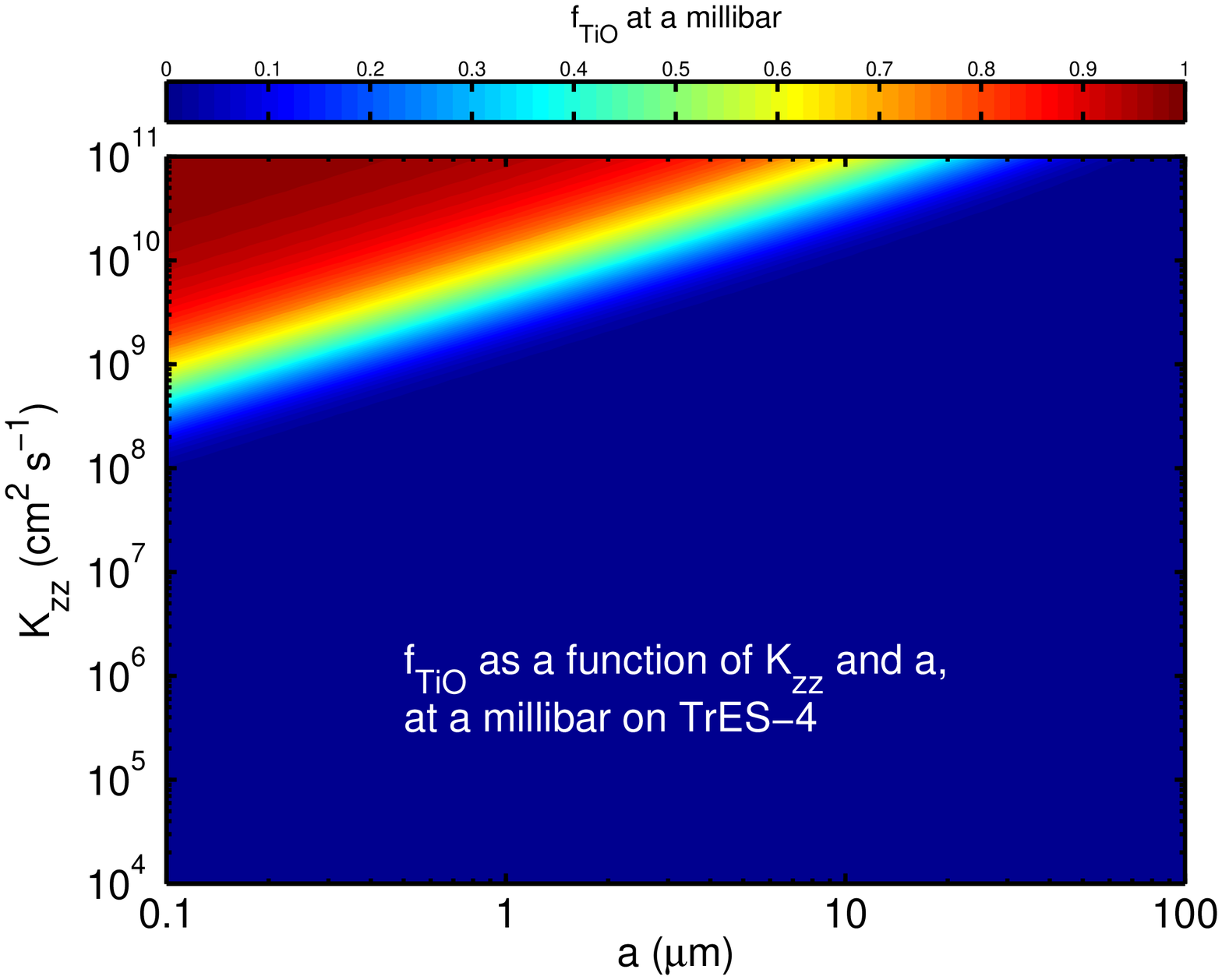}
{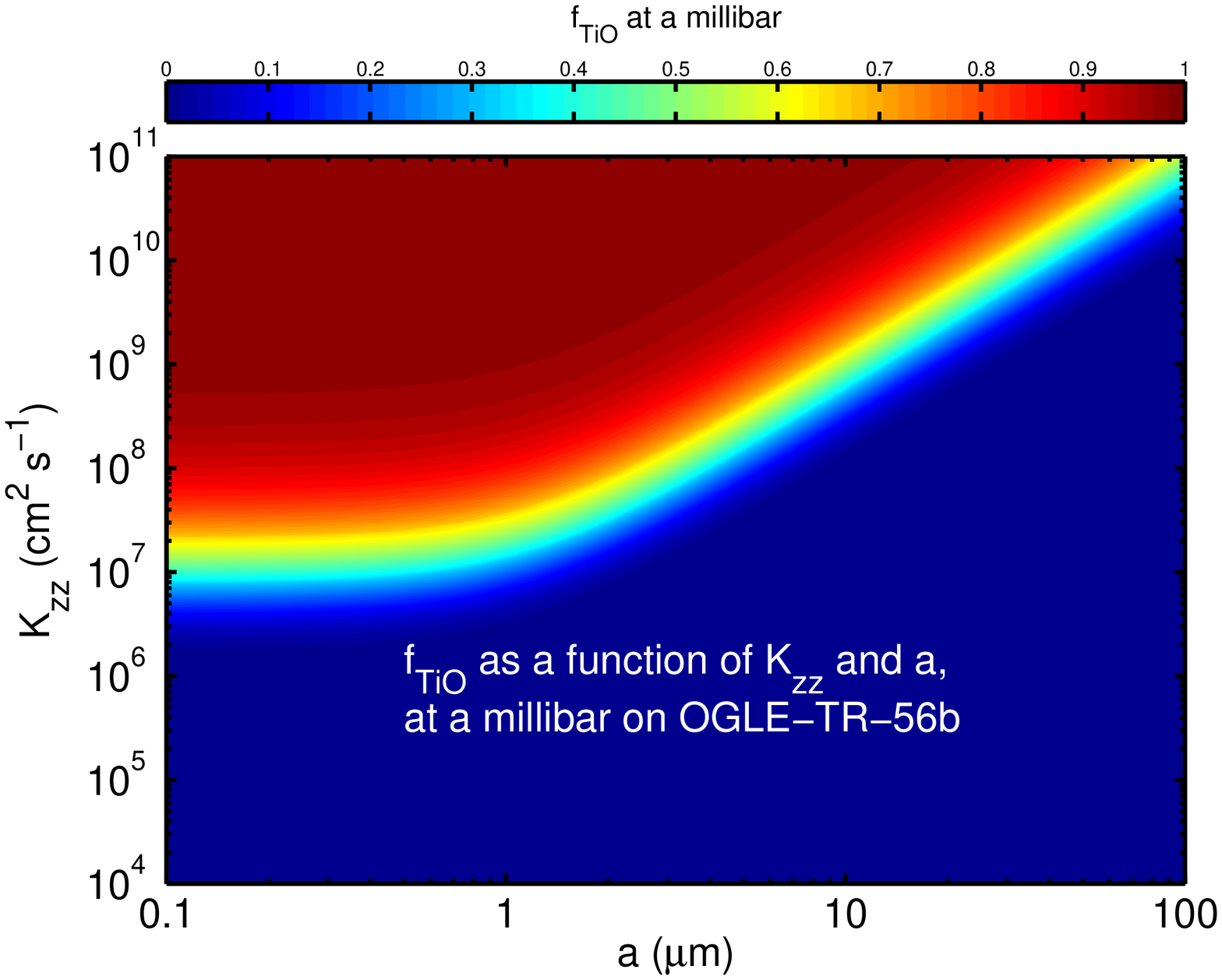}
\caption{Color-map of $f_{\rm TiO}$ at a millibar, as a function of
$a$ and $K_{zz}$ on the four planets with day-side cold traps
(HD~209458b: top-left; HD~149026b: top-right; TrES-4: bottom-left;
OGLE-TR-56b: bottom-right).  The color contours indicate the mixing
ratio of TiO at $10^{-3}$~bars relative to the interior mixing ratio
of titanium, for various combinations of particle size
(0.1-100~$\mu$m) and turbulent diffusion coefficient ($10^4$-$10^{11}
\rm~cm^2~s^{-1}$).  The green band indicates the combinations required
to achieve the fiducial value of $f_{\rm TiO} \sim 0.5$. }
\label{fig:f_at_mb_pretty}
\end{figure}
\end{landscape}
\newpage
\newpage

\end{document}